\documentclass{aa}

\usepackage[varg]{txfonts}
\usepackage{longtable}
\usepackage{bm}

\newcommand\kms{km~s$^{-1}$}
\newcommand\msun{$M_\odot$}
\newcommand\mhi{$M_{\mathrm{HI}}$}
\newcommand\mstar{$M_{*}$}

\def\be{\begin{equation}}
\def\ee{\end{equation}}
\def\a40{$\alpha$.40}
\def\arcmin{$^{\prime}$}
\def\arcsec{$^{\prime\prime}$}
\def\dg{$^{\circ}$}

\newcommand{\hi}{H\,{\sc i}}

\usepackage{natbib,twoopt}
\usepackage[breaklinks=true]{hyperref} 
\bibpunct{(}{)}{;}{a}{}{,} 
\makeatletter
\newcommandtwoopt{\citeads}[3][][]{\href{http://adsabs.harvard.edu/abs/#3}%
{\def\hyper@linkstart##1##2{}%
\let\hyper@linkend\@empty\citealp[#1][#2]{#3}}}
\newcommandtwoopt{\citepads}[3][][]{\href{http://adsabs.harvard.edu/abs/#3}%
{\def\hyper@linkstart##1##2{}%
\let\hyper@linkend\@empty\citep[#1][#2]{#3}}}
\newcommandtwoopt{\citetads}[3][][]{\href{http://adsabs.harvard.edu/abs/#3}%
{\def\hyper@linkstart##1##2{}%
\let\hyper@linkend\@empty\citet[#1][#2]{#3}}}
\newcommandtwoopt{\citeyearads}[3][][]%
{\href{http://adsabs.harvard.edu/abs/#3}
{\def\hyper@linkstart##1##2{}%
\let\hyper@linkend\@empty\citeyear[#1][#2]{#3}}}
\makeatother

\begin{document}

\title{Deep neutral hydrogen observations of Leo T with the Westerbork Synthesis Radio Telescope}
\titlerunning{Deep \hi\ observations of Leo T with WSRT}
\author{Elizabeth A. K. Adams
		\inst{1,2} 
		\and 
		Tom A. Oosterloo
		\inst{1,2}}
\institute{
ASTRON, Netherlands Institute for Radio Astronomy, Postbus 2, 7900 AA Dwingeloo, The Netherlands
 		\email{adams@astron.nl}
		\and
		Kapteyn Astronomical Institute, University of Groningen Postbus 800, 9700 AV Groningen, the Netherlands
}


\abstract{
Leo T is the lowest mass gas-rich galaxy currently known
and studies of its gas content help us understand how such marginal
galaxies survive and form stars. 
We present deep neutral hydrogen (\hi)
observations from the Westerbork Synthesis Radio Telescope in order to understand its 
\hi\ distribution and potential for 
star formation.
We find a larger \hi\ line flux than the previously accepted value, 
resulting in a 50\% larger \hi\ mass of $4.1 \times 10^5$ \msun.
The additional \hi\ flux is from low surface brightness emission that was previously missed;
with careful masking this emission can be recovered even in shallower data.
We perform a Gaussian spectral decomposition to find a cool neutral medium component (CNM)
with a mass of $3.7 \times 10^4$ \msun, or  almost 10\% of the total \hi\ mass. 
Leo T has no \hi\ emission extending from the main \hi\ body, but 
there is evidence of interaction with the Milky Way circumgalactic medium in both a
potential truncation of the \hi\ body
 and the offset of the peak \hi\ distribution from the optical center.
The CNM component of Leo T is large when compared to other dwarf galaxies,
{even though Leo T is not currently forming stars and
has a lower star formation efficiency than other gas-rich dwarf galaxies.} 
However, the \hi\ column density 
associated with the CNM component in Leo T is low.
{One possible explanation is the large CNM component
is not related to star formation potential but rather
a recent, transient phenomenon related to the interaction
of Leo T with the Milky Way {circumgalactic medium}.
}
}

\keywords{galaxies: dwarf --- 
 galaxies: ISM ---
Local Group --- 
galaxies: individual: Leo T --- galaxies: star formation
--- radio lines: galaxies}

\titlerunning{Deep \hi\ observations of Leo T with WSRT}

\maketitle{}



\section{Introduction}

Dwarf galaxies provide important insights into the processes that control and govern
star formation.
These low mass galaxies are extremely susceptible to feedback and disruption
from the local environment,
and thus serve as excellent laboratories to study these processes that are
critical to controlling star formation. Leo T is an excellent example of this;
located just 
outside the Milky Way's virial radius at a distance of 420 kpc
\citepads{2007ApJ...656L..13I,2008ApJ...680.1112D,2012ApJ...748...88W,2012ApJ...756..108C},
this galaxy has a substantial reservoir of neutral hydrogen (\hi),
$2.8 \times 10^5$ \msun\ \citepads{2008MNRAS.384..535R}, 
compared to its stellar mass, $\sim 2 \times 10^5$ \msun\ \citepads{2008ApJ...680.1112D,2012ApJ...748...88W}.
Whether it has retained this reservoir or recently reacquired it \citepads[][]{2009MNRAS.392L..45R},
it is a remarkably fragile system.
Yet, despite its proximity to the Milky Way, it shows no evidence
of tidal disruption in either its stellar population or \hi\ content \citepads{2008ApJ...680.1112D,2008MNRAS.384..535R}.

The original ground-based data for Leo T revealed a young stellar population indicating recent star 
formation; indeed Leo T is the lowest luminosity galaxy with recent star formation 
\citepads{2007ApJ...656L..13I,2008ApJ...680.1112D}.
Subsequent data with the Hubble Space Telescope showed
that Leo T had a relatively constant low star formation rate of $5 \times 10^{-5}$ \msun\ yr$^{-1}$ over the last $\sim$8 Gyr
until 25 Myr ago. At that point, the star formation history shows a drop that could
be a truncation of star formation or a low star formation rate combined with stochastic
sampling of the initial mass function resulting in a lack of high mass stars 
\citepads{2012ApJ...748...88W}.

Here, we present deep \hi\ observations of Leo T undertaken with the Westerbork 
Synthesis Radio Telescope (WSRT). These observations are a factor of three deeper
than those presented in \citetads{2008MNRAS.384..535R}. 
This improvement in sensitivity allows us to trace the lower column density material in the
outer regions of Leo T at high resolution, so that we may
 understand the distribution of \hi\ in
this extreme galaxy and its relation to the (lack of) star formation.
In Section \ref{sec:data} we present the WSRT data, and  
Section \ref{sec:hiprops} summarizes the \hi\ properties.
In Section \ref{sec:discuss} we discuss the nature of Leo T, finding
that it has a large amount of cold \hi, and postulate that this is related
to its interaction with the circumgalactic medium of the Milky Way.
We summarize our findings in Section \ref{sec:concl}.
As we find a larger line flux than the previously accepted value,
in
Appendix \ref{app:flux} we undertake an exhaustive discussion
of the \hi\ line flux determination of Leo T,
demonstrating that careful masking is necessary for recovering emission.


\section{Data}\label{sec:data}
Leo T was observed in ten sessions over the period 20 January -- 11 February 2008 with WSRT. 
Each observation was a standard 12-hr synthesis track bracketed by half an hour on a standard calibrator.
The spectral setup was 1024 channels over a 2.5 MHz bandwidth, giving a spectral
resolution of 2.44 kHz, or 0.52 \kms.
The data were edited manually to remove data affected by shadowing or radio
frequency interference (RFI). Standard calibration procedure was done in {\tt Miriad},
including correcting for $T_{sys}$ and applying bandpass and gain solutions from the calibrators
\citepads{1995ASPC...77..433S}.
Following standard WSRT practice, 
final gain solutions were dervied
from self-calibration on a continuum image
of the target field. A prototype of the Apertif pipeline, Apercal, was used to perform a phase-only self-calibration.
The continuum sources were then removed in the uv-plane using the task {\tt uvlin}.
The data were imaged in {\tt Miriad} at full spectral resolution with a robust weighting of 0.4,
resulting in a restoring beam of 57.3\arcsec $\times$ 15.7\arcsec 
(117 $\times 32.0$ pc for a distance of 420 kpc)
with a position angle of 0.1\dg. {The details of our cleaning procedure are described
in the following paragraph.}
The noise in this final data cube is 0.87 mJy beam$^{-1}$ for the 0.52 \kms\ channel.
This corresponds to a column density rms value of $5.1 \times 10^{18}$ atoms cm$^{-2}$ for 
a velocity width of 43.28 \kms, the total velocity range of Leo T\footnote{The noise in our final integrated \hi\ map over a velocity range
of 43.28 \kms\ is slightly higher: $5.4 \times 10^{18}$ atoms cm$^{-2}$. This is because the final
map includes channels with Galactic \hi\ emission; due to the lack of zero-spacing information there are
large scale features that elevate the noise slightly.}.

\begin{figure}
\centering
\includegraphics[width=\linewidth,keepaspectratio,clip=True,trim=0cm 2cm 0cm 2cm]{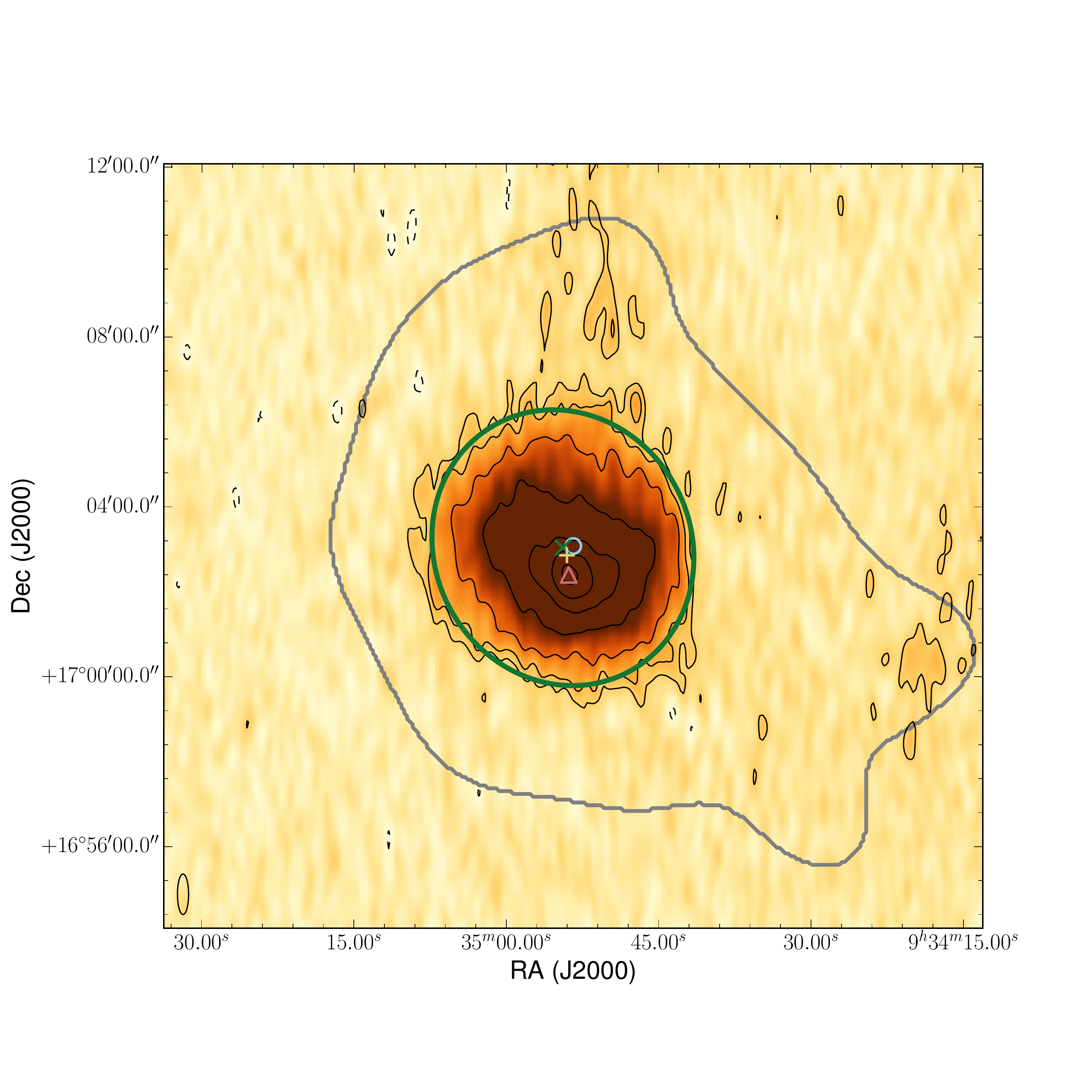}
\caption{Non-primary-beam-corrected total intensity \hi\ map of Leo T. Contours are at 
[-3, 3, 5,10, 20, 40, 60, 80]-$\sigma$; the lowest contour level is approximately $1.6 \times 10^{19}$
atoms cm$^{-2}$.
The mask used to define the Leo T source extent for cleaning is shown in gray.
The light-blue circle is the optical center, the green x the center of the \hi\ ellipse,
the yellow cross the center of the \hi\ distribution, the pink triangle is the
peak of the \hi\ distribution, {and the large green circle is the \hi\ extent at the
$2.7 \times 10^{19}$ atoms cm$^{-2}$ level}
 (Section \ref{sec:hidist}). 
}
\label{fig:mom0}
\end{figure}

As {the \hi\ velocity gradient across Leo T is small}  \citepads{2008MNRAS.384..535R},
we followed the methodology of \citetads{2013A&A...555L...7O} and \citetads{2016A&A...596A.117A} and created
 a {frequency-independent clean mask} that was 
applied to all channels with emission.
This ensured that 
low surface brightness emission 
{could be included in the total intensity map with the correct
beam area};
as discussed in Appendix \ref{app:flux}, this emission is a significant contribution to the 
total line flux.
The case of Leo T is complicated by the presence of strong foreground Galactic \hi\ emission.
Thus, we identified channels that appeared to be free of Galactic \hi\ emission but contained emission
from Leo T in order to create a single-channel image that only contains emission from Leo T.
Our Leo T-only image was centered at 49.98 \kms\ 
with a width of 14.43 \kms.
This image was created with a 60\arcsec\ taper to aid in identifying extended low surface brightness emission.
Iteratively,
 this single channel image was smoothed further
to a circular 160\arcsec\ beam,
 clipped to contain real emission,
cleaned at 60\arcsec\ resolution, smoothed and clipped to contain more emission, and so on, until
the smoothed image was clipped at the 2-$\sigma$ level to define the final mask
(shown in Figure \ref{fig:mom0}).
In addition, the Galactic foreground emission
needed to be cleaned in the channels where it coincides with {emission} from Leo T
 {because it is bright enough that the 
sidelobes from Galactic emission overlap with emission from
 Leo T}.
The Galactic emission is often narrow in velocity extent and varies significantly
from channel to channel. Thus, a second channel-based mask to clean the Galactic emission 
 was defined 
 in the same manner as above, except that the smoothed cube (not single-channel image)
 was clipped at the 3.5-$\sigma$ level to define the final channel-based mask.

\begin{figure*}
\centering
\includegraphics[width=\linewidth,keepaspectratio,clip=True, trim=2.5cm 1cm 1cm 2cm]{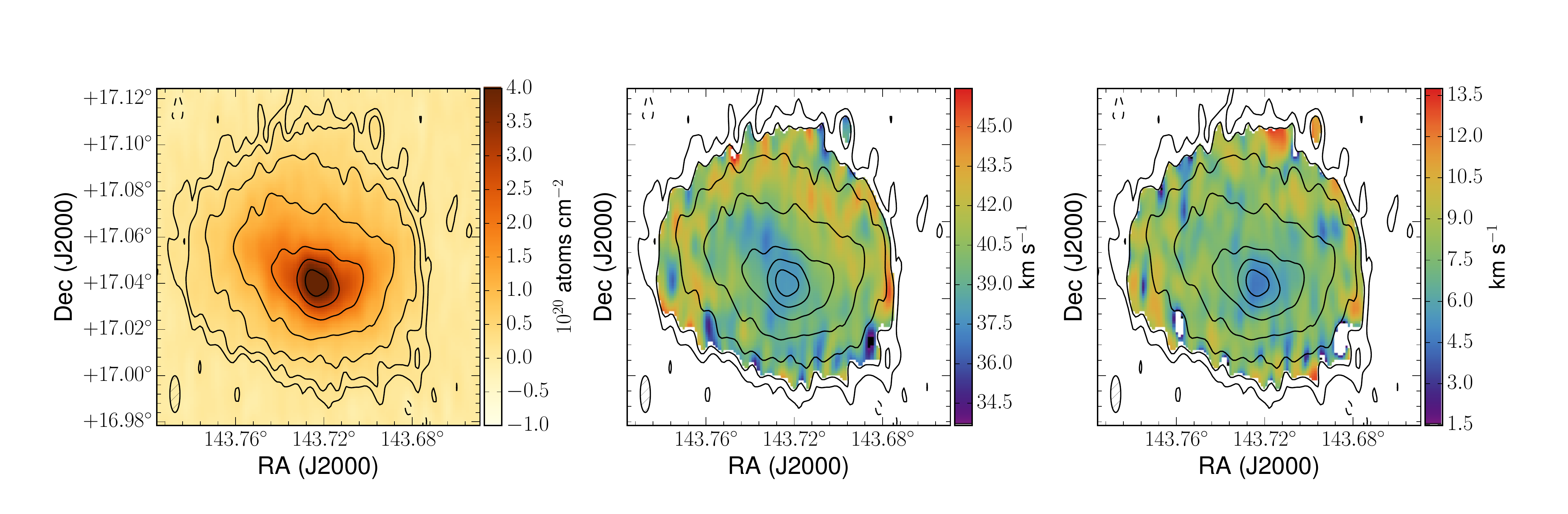}
\caption{Total intensity (primary-beam corrected), moment one (velocity field) and moment two (velocity dispersion) maps of Leo T.
\hi\ contours shown in each panel are [-1.5, 1.5, 2.5, 5, 10, 20, 30, 40] $\times 10^{19}$ atoms cm$^{-2}$;
the lowest contour level is slightly below the 3-$\sigma$ level. 
}
\label{fig:moms}
\end{figure*}

\begin{figure}
\centering
\includegraphics[keepaspectratio,width=\linewidth]{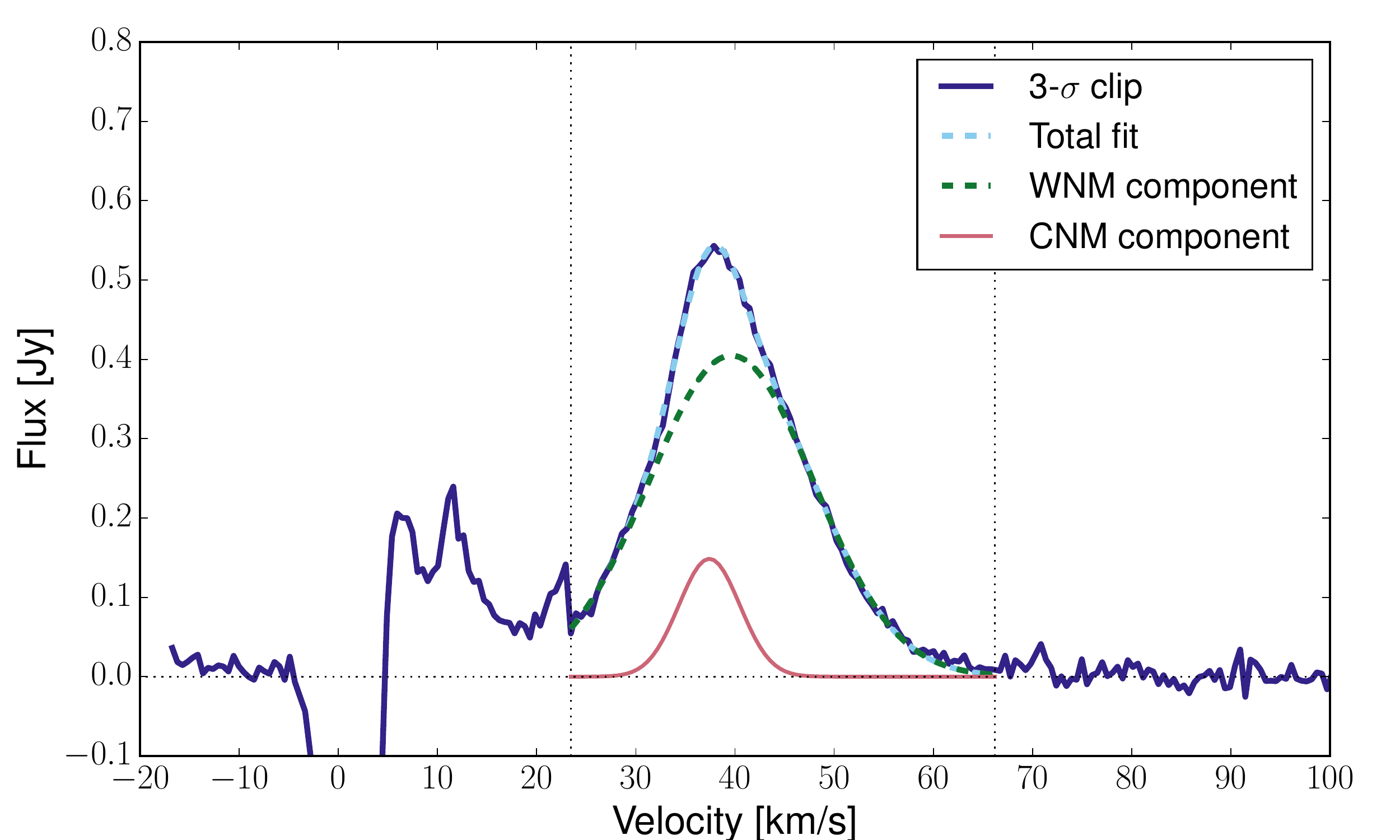}
\caption{Spectrum of Leo T based on a 3-sigma clip of the total intensity \hi\ map. 
The velocity range used for constructing the total intensity \hi\ map
is shown by the dotted vertical lines. In addition, a two Gaussian component fit to this spectrum in shown.
}
\label{fig:spec}
\end{figure}

The Leo T {frequency-independent clean} mask was applied to all channels deemed to have emission from 
Leo T ($23.45 - 66.21$ \kms), and the channel-based Galactic \hi\ mask
was kept for the same channel range (cleaning Galactic \hi\ only where Leo T emission
is also present).
Within this combined mask, a deep clean (to 0.5 times the rms) at 
full spatial resolution was performed;
the deep cleaning minimizes the impact of residual flux.
After cleaning, moment zero, one and two maps were created over
the velocity range 23.45--66.21 \kms, representing the total \hi\ intensity,
velocity field and velocity dispersion.
{We emphasize that these moment maps were created without any channel-based masking,
analogous to our cleaning strategy. This is not the standard strategy but was well
justified in the case of Leo T, which has a very small \hi\ velocity gradient. Appendix \ref{app:flux} 
contains a discussion on how this strategy aids in flux recovery.}
These maps are presented in Figure \ref{fig:moms},
where the total intensity \hi\ map is primary-beam corrected and in units of 
column density, assuming optically thin emission.
Figure \ref{fig:spec} presents a global spectrum, extracted by clipping the (non-primary-beam-corrected)
moment zero map at the 3-$\sigma$ level, neglecting the potential northern
and western extensions of emission (see Section \ref{sec:exten}),
and applying that as a mask to the primary-beam-corrected data cube.

We used the total intensity \hi\ map and the global spectrum to measure
the line flux of Leo T.
Integrating the spectrum over the velocity range used to create the total intensity \hi\ map
resulted in a line flux of 9.4 Jy \kms.
{As can be seen in Figure \ref{fig:spec}, the velocity range used
to create the total intensity map misses some emission from Leo T that is spectrally
confused with the bright Galactic foreground. Thus,
}
we also calculated the line flux by fitting Gaussian components to the spectrum. 
As discussed below in Section \ref{sec:gaussdecomp}, two Gaussian components fit the 
spectrum significantly better than a single Gaussian component.
This fit is shown in Figure \ref{fig:spec}; 
integrating it resulted in a line flux of 9.6 Jy \kms.
These two line flux values are significantly higher than previously reported in the literature, either based on single-dish
HIPASS data \citepads{2007ApJ...656L..13I} or shallower WSRT observations \citepads{2008MNRAS.384..535R}, although consistent with a second measurement
based on the HIPASS data \citepads{2009ApJ...696..385G}.
In Appendix \ref{app:flux} we exhaustively discuss the flux determination of Leo T; 
we compare to previous work and single-dish measurements and explore different
methods of deriving the flux. The crucial difference between this work and previous work is
that the {frequency-independent clean mask allowed  creation of a total intensity
\hi\ map without channel-based masking, which included} low intensity emission 
{from individual channels}
 that would otherwise be excluded from 
a channel-based mask.
When using a mask based on smoothing the data, we found a final flux value of $9.9 \pm 1.0$ Jy \kms\
{from fitting the integrated spectrum with two Gaussian components};
the uncertainty encompasses different methods of determining the line flux and defining 
the extent of Leo T {when using a frequency-independent mask}, 
plus a systematic uncertainty of $\sim$10\% in the calibration. 
We also note that our peak column density value ($4.6 \times 10^{20}$ atoms cm$^{-2}$) is $\sim$15\%
lower than that of \citetads[][$5.3 \times 10^{20}$ atoms cm$^{-2}$]{2008MNRAS.384..535R}; 
however, our restoring beam is $\sim$40\% larger in area than that work so we smoothed the structure out slightly,
accounting for our lower peak column density value.

\section{\hi\ properties of Leo T}\label{sec:hiprops}

\subsection{\hi\ mass and gas fraction}

With our updated line flux value, the \hi\ mass associated with Leo T also increases,
to $4.1  \times 10^{5}$ \msun\ (for a distance of 420 kpc),
50\%  higher than \hi\ mass previously associated with Leo T.
Despite the larger gas mass, we find a similar gas fraction as \citetads{2008MNRAS.384..535R}
due to an updated stellar mass that is 
{increased by a similar amount}.
Using resolved stellar observations of Leo T with the {\it HST}, \citetads{2012ApJ...748...88W} report
a stellar mass of $1.05^{+0.27}_{-0.23} \times 10^5$ \msun\ over their field of view, which corresponds to 
the area of Leo T within the half-light radius. Thus, we adopt $2 \times 10^5$ \msun\ as our stellar mass 
of Leo T.
This is consistent with the mass-to-light ratio of two used by \citetads{2008MNRAS.384..535R} when the luminosity of Leo T is
updated for the deeper photometry of \citetads{2008ApJ...680.1112D}.
Accounting for the presence of helium in the atomic gas ($M_{gas} = 1.33\, \times$ \mhi),
we find a neutral gas fraction, $f_{gas} = M_{gas}/(M_{gas}+$\mstar), of 0.73.

\subsection{\hi\ kinematics}\label{sec:hikin}

\begin{figure*}
\centering
\includegraphics[keepaspectratio,width=0.45\linewidth,clip=True, trim=0cm 2cm 0cm 0.25cm]{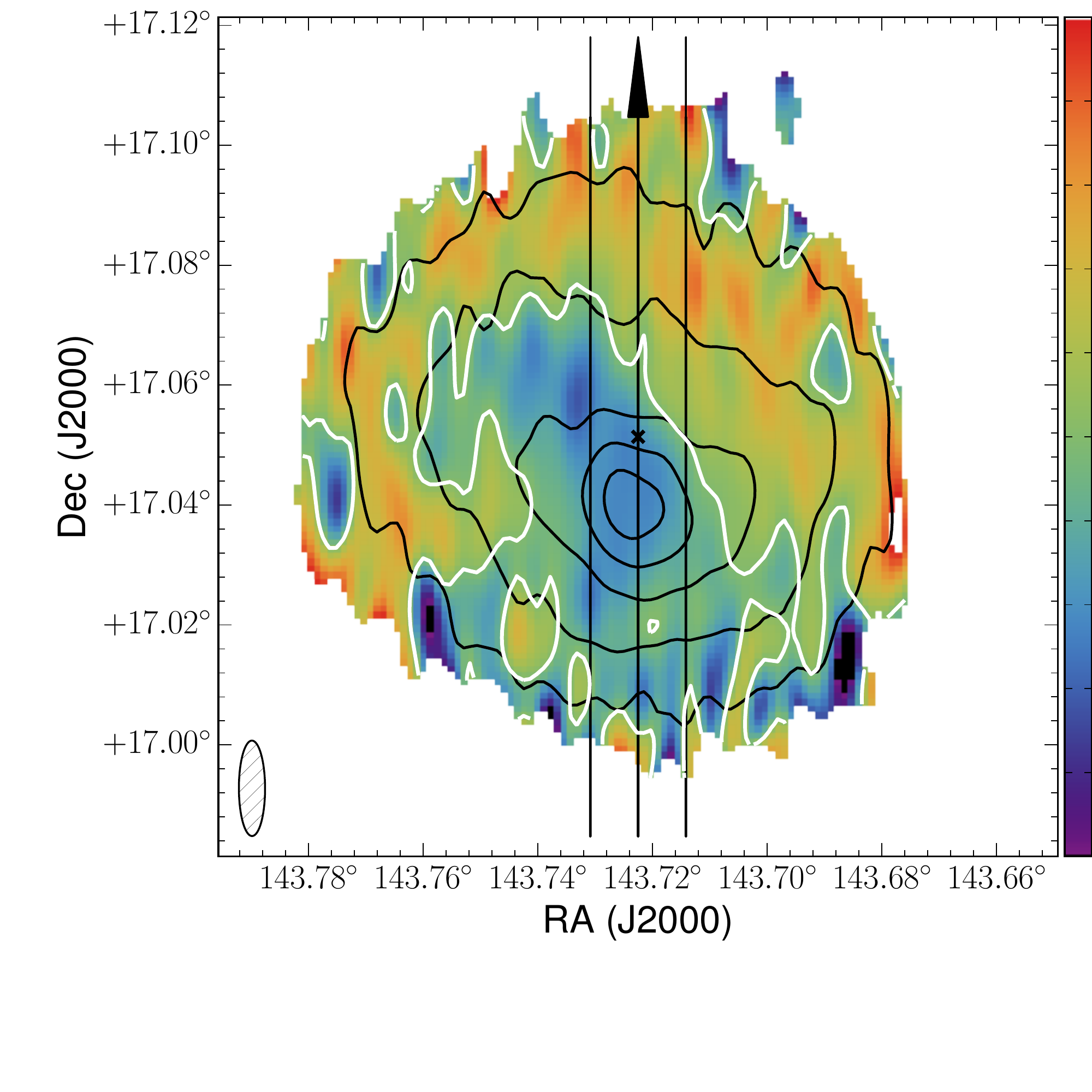}
\quad
\includegraphics[keepaspectratio,width=0.45\linewidth]{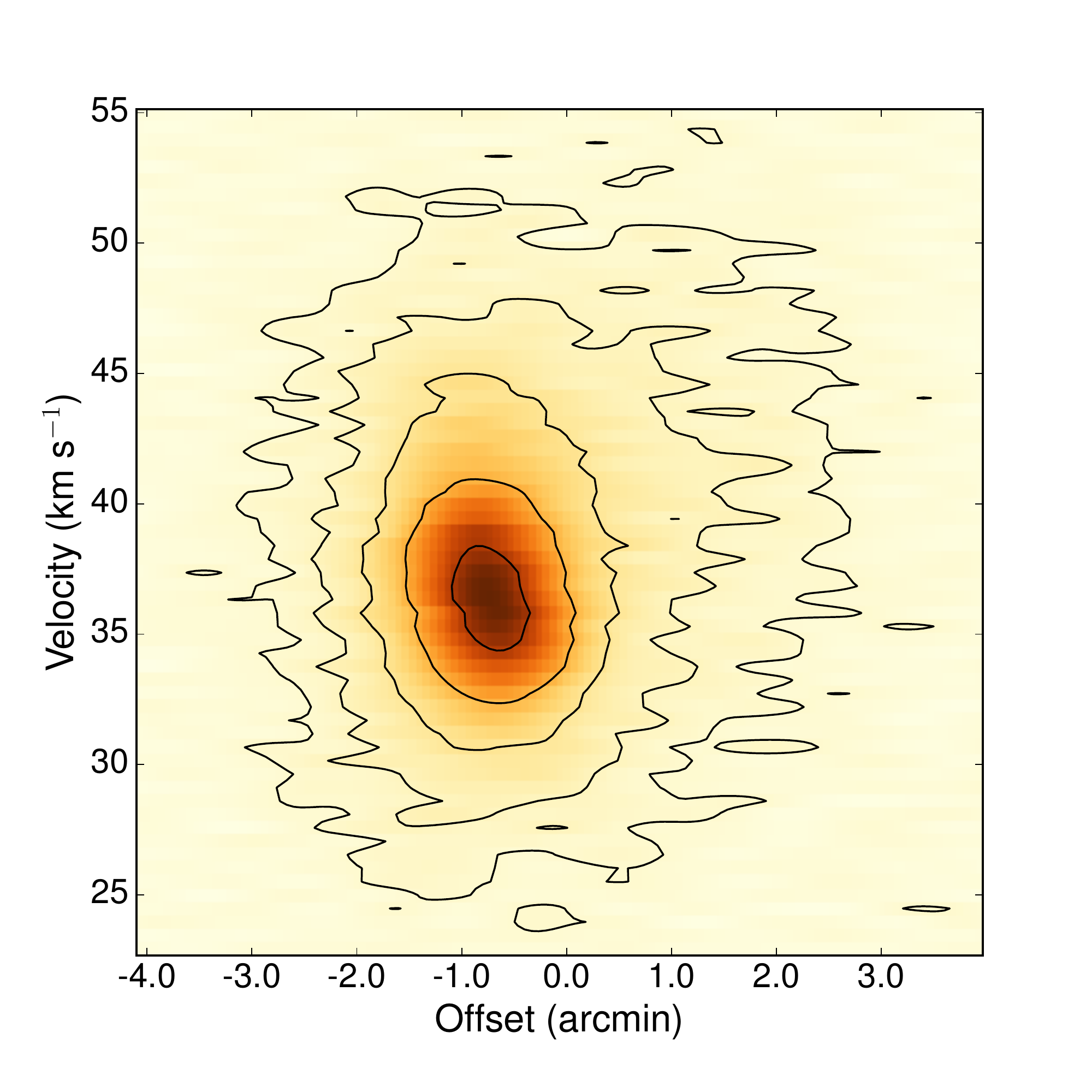}
\caption{
{
{\it Left:} Velocity field of Leo T as derived from the moment one map with the isovelocity contour at
40 \kms\ highlighted in white. The \hi\ column density contours at [0.5, 1, 2, 3, 4] $\times 10^{20}$ atoms cm$^{-2}$
are also shown in black. 
{\it Right:}
Position-velocity slice of the \hi\ in Leo T along a position angle of zero degrees with a width of 60\arcsec\ and centered
on the optical centroid of the galaxy. The
angle of this slice,  extent, and center are shown overlaid on the velocity field in the left panel. The contours
are at [2, 4, 8, 16, 32] times the rms of the data cube.
}
}
\label{fig:pv}
\end{figure*}

{ 
The velocity field (moment one map) presented in
Figure \ref{fig:moms} potentially shows a small velocity gradient
 across Leo T from north to south. In Figure \ref{fig:pv} we explore this potential
gradient further. 
The left panel shows the velocity field with the isovelocity contour of 40 \kms\
highlighted. There is a clear asymmetry between the north and south with
higher velocity gas to the north and lower velocity gas to the south.
The right panel shows a position-velocity slice extracted along a position angle of 
zero degrees with a width of 60\arcsec\ using the task {\tt impv} in {\tt CASA}.
This position-velocity slice shows no evidence for a large-scale velocity gradient across
the extent of Leo T. 
This implies that either Leo T is not rotationally supported, or that it is essentially completely
face-on so that there is no rotation velocity along the line-of-sight.
In Section \ref{sec:gaussdecomp}, we determine that Leo T has two interstellar medium components:
a cool and warm neutral medium. These two components are offset from each spatially and in velocity;
these offsets between the two components could be responsible for the appearance of a gradient
in the velocity field.
}

\subsection{The state of the ISM}\label{sec:gaussdecomp}

A common technique for understanding the interstellar medium (ISM) in dwarf 
galaxies is to decompose
{spectra, either integrated or spatially resolved,}
 into two Gaussian components: a narrow Gaussian component
($\sigma \lesssim 5$ \kms) and a broad Gaussian component ($\sigma \gtrsim 10$ \kms).
These components are interpreted as corresponding to a cool neutral medium (CNM; $T \lesssim 1000$ K)
and a warm neutral medium (WNM; $T \gtrsim 5000$ K) component of the \hi\ gas
\citepads[e.g.,][]{1996ApJ...462..203Y,1997ApJ...476..127Y,1997ApJ...490..710Y,
2003ApJ...592..111Y, 2006AJ....131..363D,2012ApJ...757...84W}.
{
The previous work by \citetads{2008MNRAS.384..535R} with shallower WSRT data showed
that Leo T contains a CNM component.
With our deeper data, we expanded on this work to more fully characterize the CNM in Leo T.

First, as Leo T has {essentially no \hi\ velocity gradient} (Section \ref{sec:hikin}), we used the global spectrum
based on a 3-$\sigma$ direct clip
(shown in Figure \ref{fig:spec}) to place limits
on the CNM component of Leo T.
We used the full velocity range of 23.45--66.21 \kms\ for the decomposition
of the spectrum,
and fit both a single and double Gaussian component.
The two component Gaussian fit is strongly preferred to a single Gaussian component.
The $\chi^2_{red}$ value for the two component fit is more than an order of magnitude smaller
than that for the single component. According to a single-tailed $F$-test the probability
of improvement for two Gaussian components over a single Gaussian component is
essentially 100\%.
The global dispersion of the CNM component is 3.1 \kms, and it has a line flux
of $1.1 \pm 0.1$ Jy \kms, or an \hi\ mass
of $4.6 \times 10^4$ \msun. The global dispersion of the WNM component is 8.3 \kms,
and it has a line flux of $8.4 \pm 0.8$ Jy \kms, or an \hi\ mass of $3.5 \times 10^5$ \msun.
In Appendix \ref{app:flux}, we present the decomposition for the global spectrum from smoothed data
(used for the final total line flux value). The results are similar: a CNM line flux of $1.1 \pm 0.2$ Jy \kms,
a larger WNM flux of $8.7 \pm 0.9$ Jy \kms, corresponding to the larger total
line flux derived from this spectrum,
{and the same velocity dispersion values within the errors}.

We also undertook a spatial analysis of the CNM component of Leo T,
following a similar methodology to \citetads{2012ApJ...757...84W}.
We considered the velocity range 23.45 -- 106.40 \kms; the lower 
{end of the} velocity 
range is the cutoff of Leo T emission before confusion with Galactic \hi,
and the upper end is the edge of our data cube.
We fit the spectra at each pixel with a single and double Gaussian
using the python package {\tt lmfit} \citep{newville_2014_11813}.
We implemented a signal-to-noise ratio (S/N) cutoff for the spectra being fit,
requiring the peak value be at least ten times the rms. We also
required a minimum S/N of 3.1 for each component of the double Gaussian fit.
In order to avoid fitting noise peaks, we 
 required the minimum velocity dispersion to be 1 \kms,
 twice the velocity resolution of our data cube. 
Finally, we only accepted a double Gaussian fit if the probability it improved
the fit compared to a single Gaussian was 95\% or greater according to a single-tailed $F$-test.
The results of our spatial decomposition of the ISM of Leo T
are shown in Figure \ref{fig:decomp}.
The top row is the narrow Gaussian component, representing the CNM,  in regions where the double Gaussian fit is preferred.
The bottom row is the combination of the broad Gaussian component from regions where the
double Gaussian fit is preferred and the single Gaussian fit in regions where that is preferred.
Combined, this represents the WNM component as 
the single Gaussian fit always has a a velocity dispersion $\gtrsim 6$ \kms,
implying it corresponds to the WNM.

As evident in Figure \ref{fig:decomp}, the CNM component coincides 
precisely with the peak \hi\ distribution, as
expected since that is the densest material.
Indeed, above the $N_{HI}$ level of $3 \times 10^{20}$
atoms cm$^{-2}$, a CNM component is always present,
consistent with a threshold model for the formation
of CNM \citepads[e.g.,][]{2004ApJ...609..667S}.
Based on the spatial decomposition of Leo T,
we found a CNM line flux of 0.86 Jy \kms, slightly lower
than that found in the decomposition of the global spectrum
(1.1 Jy \kms), 
{although consistent within the errors}. 
One explanation for
this discrepancy is that low-level CNM
emission is discarded from the pixel-based decomposition,
e.g., for being below the S/N limits,
but contributes in the global decomposition.
However, as seen in Table \ref{tab:fluxvals} in the Appendix,
the CNM line flux determined from the global spectrum is sensitive
to the exact global spectrum used.
We adopt $0.9 \pm 0.2$ Jy \kms\ as
the total CNM line flux.
We then assign the rest of the total line
flux, 9.0 $\pm$ 1.0 Jy \kms, to the WNM component.

Based on the spatial decomposition,
the median velocity dispersion for the CNM is 2.5 ${\pm 0.1}$ \kms\ ($T \sim 780$ K),
and that of the WNM is 7.1 ${\pm 0.4}$ \kms\ ($T\sim 6300$ K).
{These values are smaller than those found for the global
spectrum.}
As the bulk velocity motions of the gas in Leo T can broaden
the measured velocity dispersion in the global spectrum,
we adopt 
the median velocity dispersions from the spatial decomposition
as the indicative dispersion of the two components.
Our CNM component is slightly broader ($\sim$0.5 \kms) than that measured 
by \citetads{2008MNRAS.384..535R} while our WNM component has
the same linewidth.

{We also note that the WNM and CNM components have different
central velocities, as can be seen in Figures \ref{fig:spec} and \ref{fig:decomp}.
In Table \ref{tab:hiprops}, we report the central velocities of each component
from the fit to the global spectrum. The two components have a significant velocity offset
of $\Delta v = 2$ \kms.
Understanding the origin of this velocity offset would offer insight into the physical state of
the gas in Leo T.
}

\begin{figure*}
\centering
\includegraphics[keepaspectratio,width=\linewidth,clip=True,trim=1cm 1.5cm 0cm 2cm]{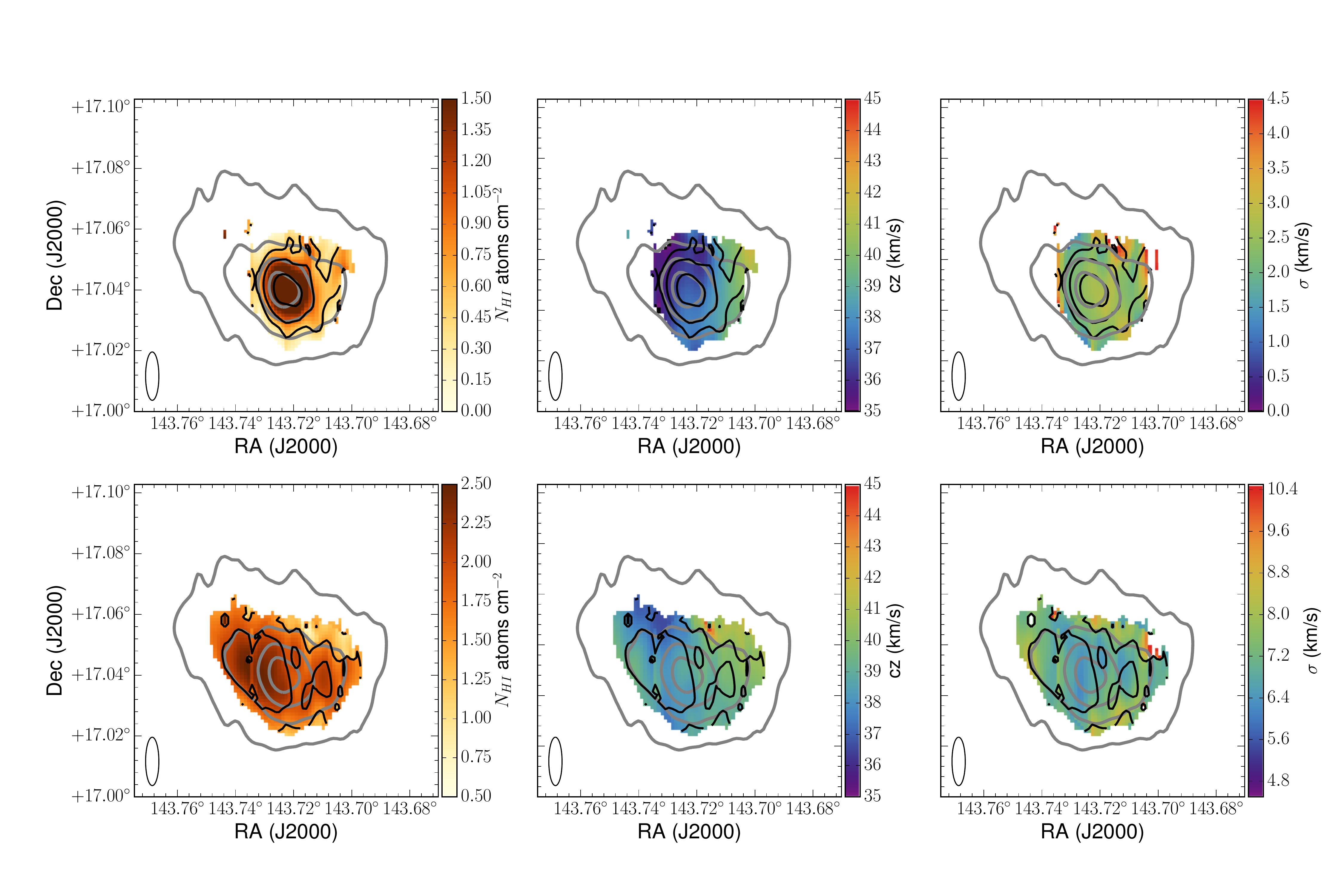}
\caption{Results of the spatial decomposition of Leo T into a WNM and CNM component.
The top row is the CNM component and the bottom the WNM component. Left to right the columns
are: column density, central velocity, and velocity dispersion of the gas. The gray contours are the
[1, 2, 3, 4] $\times 10^{20}$ atoms cm$^{-2}$ level from the total column density map in Figure \ref{fig:moms}.
The black contours are the column density for each component; for the CNM the levels are
[0.5, 1, 2] $\times 10^{20}$ atoms cm$^{-2}$ and for the WNM the levels are [1.5, 2] $\times 10^{20}$
atoms cm$^{-2}$.}
\label{fig:decomp}
\end{figure*}

\begin{table}
\caption{Properties of Leo T}
\label{tab:hiprops}
\centering
\begin{tabular}{ll}
\hline \hline
Property & Value \\
\hline
Optical center & 9$^{\mathrm{h}}$34$^{\mathrm{m}}$53.4$^{\mathrm{s}}$ +17\dg03\arcmin05\arcsec \\ 
\hi\ center & 9$^{\mathrm{h}}$34$^{\mathrm{m}}$54.0$^{\mathrm{s}}$ +17\dg02\arcmin52\arcsec \\
Distance & 420 kpc \\
$S_{\mathrm{int}}$ & 9.9 $\pm$ 1.0 Jy \kms \\
$S_{\mathrm{int, CNM}}$ & 0.9 $\pm$ 0.2 Jy \kms \\
$S_{\mathrm{int, WNM}}$ & 9.0 $\pm$ 1.0 Jy \kms \\
$\sigma_{\mathrm{CNM}}$ & 2.5 ${\pm 0.1}$ \kms \\
$\sigma_{\mathrm{WNM}}$ & 7.1 ${\pm 0.4}$ \kms \\
${\mathrm{v_{cen, CNM}}}$ & ${37.4 \pm 0.1}$ {\kms}\\
${\mathrm{v_{cen, WNM}}}$ & ${39.6 \pm 0.1}$ {\kms}\\
$N_{\mathrm{HI,peak}}$ & $4.6 \times 10^{20}$ atoms cm$^{-2}$\\
$a \times b$\tablefootmark{a} & 3.3\arcmin $\times$ 3.0\arcmin \\
$r_{\mathrm{HI}}$\tablefootmark{a} & 400 pc \\
$M_{\mathrm{HI}}$ & $4.1{\pm 0.4} \times 10^5$ \msun \\
$M_{\mathrm{CNM}}$ & $ {0.37 \pm 0.08 \times 10^5}$ \msun\\
$M_{\mathrm{WNM}}$ & $3.7 {\pm 0.4} \times 10^5$ \msun \\
$M_{\star}$ & $2.0 \times 10^5$ \msun \\
$M_{\mathrm{dyn}}$\tablefootmark{a} & $1.9 \times 10^7$ \msun \\
$f_{\mathrm{gas}}$ & 0.73 \\
\hline
\end{tabular}
\tablefoot{
\tablefootmark{a}{\hi\ extent is measured at the $2.7 \times 10^{19}$ atoms cm$^{-2}$ level.}
}
\end{table}




\subsection{\hi\ Distribution}\label{sec:hidist}

In order to understand the distribution of gas in the main \hi\ disk,
we undertook a moments analysis, 
 following \citetads{1995MNRAS.272..821B} and using
the primary-beam-corrected total intensity \hi\ map clipped at the $2.7 \times 10^{19}$ atoms cm$^{-2}$ level
(approximately 5-$\sigma$).
Using second order moments, we fit an ellipse to this extent,
shown in Figure \ref{fig:mom0}.
This ellipse has
semi-major axes of 3.3\arcmin $\times$ 3.0\arcmin, with a position angle of 28\dg,
corresponding
to an almost circular distribution of \hi\ -- the ellipticity is 0.1. 
Adopting an \hi\ radius of 3.3\arcmin\ implies that the \hi\ radius at the
 $2.7 \times 10^{19}$ atoms cm$^{-2}$ level is 400 pc.
\citetads{2008MNRAS.384..535R} trace the \hi\ extent to a slightly lower column density level 
($2 \times 10^{19}$ atoms cm$^{-2}$)
but find a smaller radius
of $\sim$300 pc due to their missing emission ({See Appendix \ref{app:flux}}).

We also used weighted first order moments to find the center of the \hi\
distribution of Leo T.
Our derived center of the \hi\ distribution is 
9$^{\mathrm{h}}$34$^{\mathrm{m}}$54.0$^{\mathrm{s}}$ +17\dg02\arcmin52\arcsec, and is shown 
in Figure \ref{fig:mom0}.
As can be seen in the figure, this is close to the 
optical center, but  offset from the peak of the \hi\ distribution,
found using {\tt maxfit} in {\tt Miriad}.

These deeper observations trace the \hi\ disk to a larger radial extent, 
allowing us to both probe more of the dark matter halo and understand the shape
of the \hi\ radial profile.
From the virial theorem, the dynamical mass within the \hi\ extent is given by
$M_{dyn} = 3 r \sigma^2 / G$
where r is the radius to which we trace the \hi\ and
$\sigma$ is the line-of-sight velocity dispersion of the gas.
We trace the \hi\ to a radius of 400 pc and measure a line-of-sight
velocity dispersion for the WNM component of 7.1 \kms, resulting in a lower
limit to the dynamical mass of $1.4 \times 10^7$ \msun.
The WNM velocity dispersion from the global spectrum is a slightly larger
value: 8.3 \kms. This may be more appropriate for the calculation of the dynamical mass as it includes
the bulk chaotic motions of the gas in addition to the thermal support. 
Then the lower limit on the dynamical mass increases to $1.9 \times 10^7$ \msun.
{As discussed in Section \ref{sec:hikin}, there is no
evidence for rotation in Leo T. If this is due to a face-on orientation, rather than intrinsic to the galaxy,
the dynamical mass could increase substantially.}

Our methodology of combining all channels without masking also allowed
us to  integrate in ellipses over the total intensity map to push beyond the direct detection limits.
Figure \ref{fig:radialprofile} presents the radial profile of \hi\ in Leo T,
found using the task {\tt ellint} in {\tt GIPSY}, centered on the optical
centroid and assuming an inclination of zero.
As determined above, the ellipticity of the \hi\ is close to zero,
and the stellar population also has an ellipticity of 
$\sim$0 \citepads{2007ApJ...656L..13I},
so we adopted circular apertures. 
The center of the \hi\ distribution is close to the optical center,
while the peak of the \hi\ distribution is offset.
The \hi\ is more sensitive to environmental processes (see Section \ref{sec:env}),
so we adopted the optical center as the best tracer of the center
of the dark matter potential.
The rings were chosen to have centers from 7.5\arcsec\--307.5\arcsec\ with a width of 15\arcsec\
(corresponding to the minor axis of the restoring beam
and $\sim 1/4$ the major axis of the restoring beam).
Since the emission seen to the north and west is likely Galactic \hi\ emission (Section \ref{sec:exten}),
we also integrated in segments of rings
corresponding to the four cardinal directions:
north (-45\dg\ -- 45\dg), east (45\dg\ -- 135\dg), south (135\dg\--225\dg), and west (225\dg\--315\dg).
The left panels of Figure \ref{fig:radialprofile} shows the total averaged radial profile,
along with the profiles for each of the four segments, for the total column density map,
as seen in the left panel of Figure \ref{fig:moms}.
We also produced a "WNM-only" column density map by subtracting
the contribution of the CNM (upper left of Figure \ref{fig:decomp}) from
the total column density map;
the \hi\ radial profiles for this map are shown in the right panels
of Figure \ref{fig:radialprofile}.

\begin{figure*}
\centering
\includegraphics[keepaspectratio,width=\linewidth,clip=True,trim=2.5cm 1.5cm 2.5cm 1.5cm]{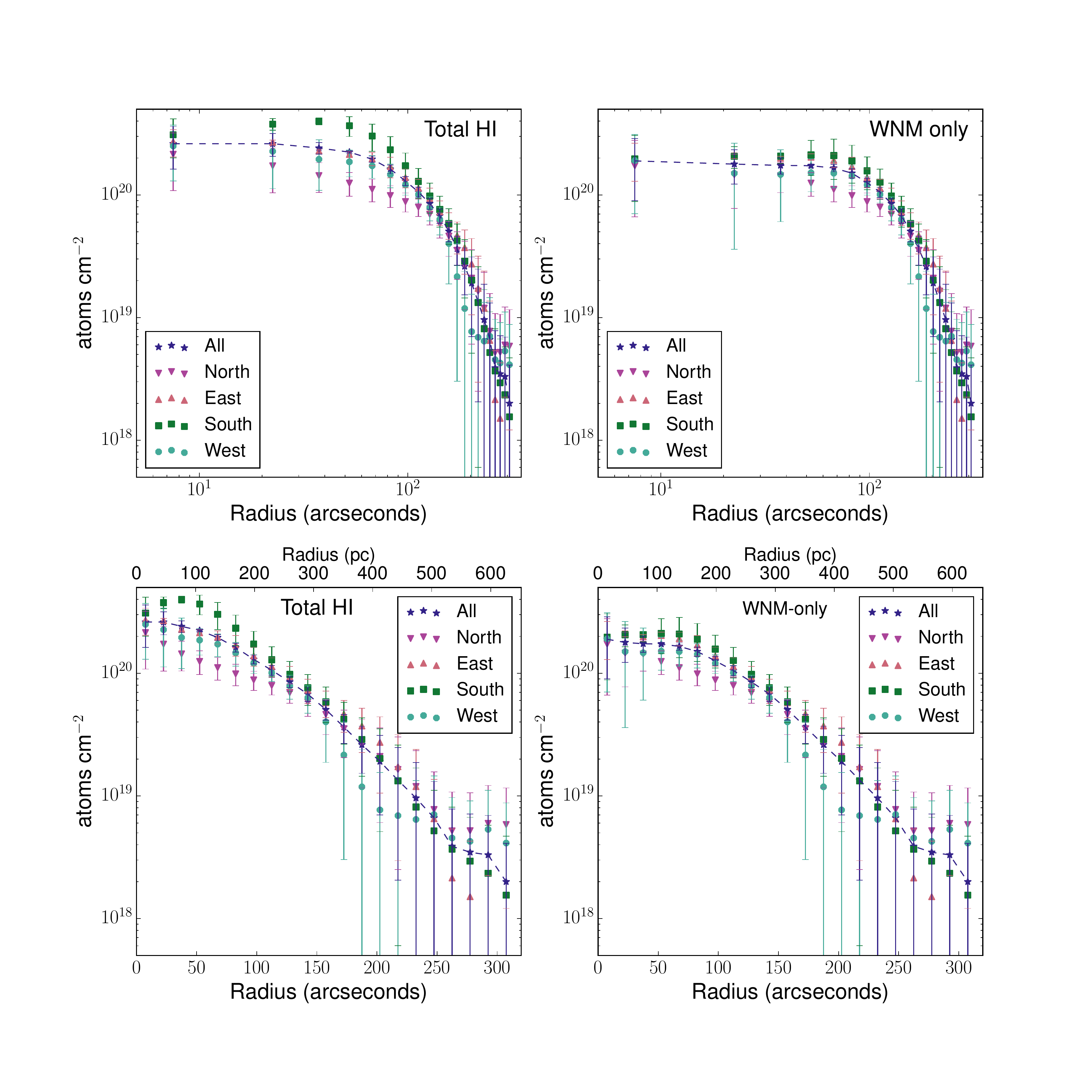}
\caption{\hi\ column density radial profile for Leo T. The profile
for the total \hi\ map is shown on the left; the right panel is for the WNM component only, after subtracting the CNM component. In addition to the global
radial profile, radial profiles for ninety-degree segments in the four cardinal directions are also shown.}
\label{fig:radialprofile}
\end{figure*}

In order to use the radial profile to trace the $N_{HI}$ distribution of Leo T beyond
the direct detection limits, the errors must be carefully accounted for. 
The statistical error comes from the noise of the \hi\ intensity map, which is
$5.4 \times 10^{18}$ atoms cm$^{-2}$. 
Systematic errors arise from multiple processes: the contribution of flux
from un-cleaned residuals, the channel-range and method used to produce the
total intensity \hi\ map, the chosen center of the radial profile, and systematics in calibration.
In order to account for systematics in creation of the total intensity \hi\ map,
we created moment zero maps over a narrower velocity range (23.45--57.45 \kms). 
In addition, 
{we also created a total intensity \hi\ map
with channel-based masking using the
\hi\ Source Finding Application, {\tt SoFiA} 
\citepads[][see Appendix \ref{app:flux} for details]{2015MNRAS.448.1922S}.
}
For both of these alternative total \hi\ intensity maps, we find the difference relative 
to our nominal profile and treat that as an error.
As the \hi\ distribution is offset from the optical center and not fully symmetric, the radial
profile is also sensitive to the center position chosen. We also produced radial profiles
using the center of the  \hi\ distribution as derived above;
the difference between this profile and the nominal profile is included as another source
of error.
The deep cleaning minimizes the impact of systematics from flux in the residuals; the 
contribution from the residuals is below all the other sources of uncertainty for all radii.
We also note that there is a $\sim$10\% systematic uncertainty from the accuracy of calibration;
this uncertainty does not change the shape of the profile but simply shifts the whole profile globally,
and so we do not include it in Figure \ref{fig:radialprofile}.
{There is an additional, small source of error ($\sim$2\%) arising
from the fact that not all Leo T emission is included
in the total intensity \hi\ map due to spectral confusion with the Galactic foreground. This error
should affect all radii equally given the lack of rotation velocity in Leo T.}
The errors from the use of a different central position dominate in the inner region of the profile.
At the outer extent (beyond $\sim$200\arcsec),
the errors are dominated by the use of a different method, the channel-based method,
to find the total \hi\ intensity map. At these large radii, the emission in many
channels is too low to be included in the source extent, and the mean surface brightness in the
{channel-based} map is suppressed.

The upper panels of Figure \ref{fig:radialprofile} show
the radial profile with a logarithmic axis for the radius.
This highlights the drop in the radial profile beyond the $\sim$100\arcsec\
 ($\sim$200 kpc) distance,
consistent with the models of \citetads{2013ApJ...777..119F} and \citetads{2017MNRAS.465.3913B}
for gas in low mass dark matter halos.
In the bottom panels, the radius is shown on a linear scale.
This highlights the fact that we do not detect the "edge" of this galaxy 
\citepads[e.g.,][]{1993ApJ...414...41M}.
While the \hi\ radial profile appears to level off at the outermost extent, the 
values are formally consistent with zero here. 
Given that the leveling off is dominated by the west and north segments, where there is 
Galactic \hi\ emission (see Section \ref{sec:exten}), the most
likely hypothesis is that this is foreground Galactic \hi\ emission 
contributing to the integrated emission.
At the $\sim$250\arcsec\ (500 pc) radius before this apparent plateau, the mean
column density value is 
$6\times 10^{18}$ atoms cm$^{-2}$;
we adopt this as the robust limit of where we can confidently trace the gas distribution.
With deeper observations, we would be able to probe 
the \hi\ distribution to fainter column densities at further radial extent. 
However, our total line flux for Leo T would not change as this emission
contributes very little to the global line flux value.
Indeed, the flux contained in emission below the $1.6 \times 10^{19}$ atoms cm$^{-2}$ level
is only $\sim$3\% of the total flux.

At the innermost radii, there is a clear asymmetry in the north and south radial profiles
for the total \hi\ map. This is expected as the peak \hi\ emission is offset to the south from the optical
center. In the WNM-only radial profile, this asymmetry is lessened, although still present. This
indicates that the bulk of the \hi\ offset from the optical center is in a CNM-phase, consistent
with the findings of Section \ref{sec:gaussdecomp}.
The western radial profile drops much more steeply than the other directions;
this matches the more closely packed \hi\ contours on the western edge of Leo T
as seen in Figure \ref{fig:mom0}.


\subsection{Potential Extended Emission}\label{sec:exten}

\begin{figure}
\centering
\includegraphics[keepaspectratio,width=\linewidth]{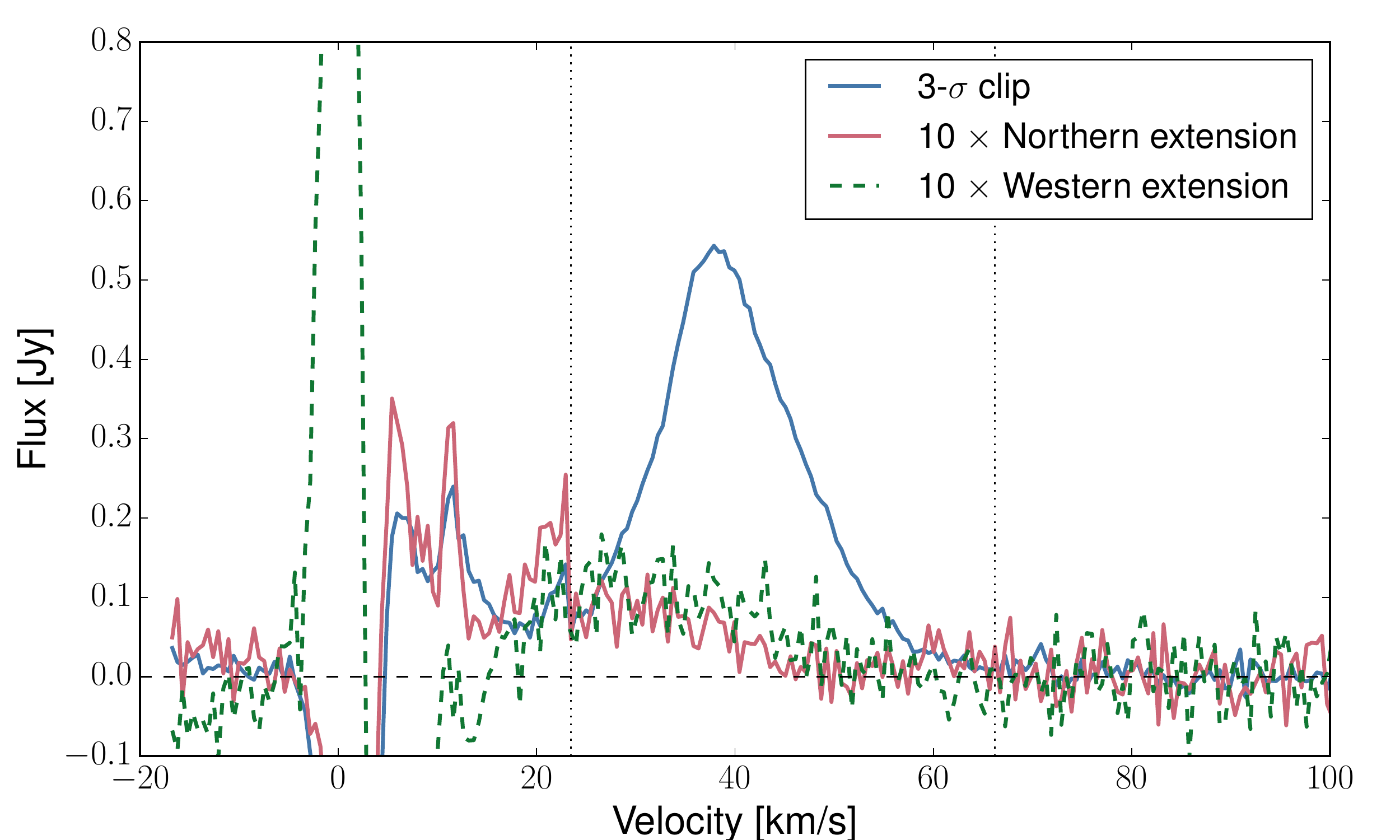}
\caption{Spectra of the potential northern and western extensions, multiplied by a factor of ten
to aid in comparison to spectrum of Leo T shown for comparison. The dashed black line indicates
the baseline level, and the dotted vertical lines the velocity extent used for creating the total 
intensity \hi\ map.}
\label{fig:specext}
\end{figure}

The total intensity \hi\ map of Leo T (Figure \ref{fig:mom0}) shows potential emission extended
to the north and west of the main body of Leo T. However, 
foreground Galactic \hi\ is present at the same velocities as Leo T,
and this extended emission may instead be associated with the Galaxy, rather than Leo T.
In order to address the nature of this potential extended emission, we smoothed the total intensity  map
with a 100\arcsec\ {Gaussian} and clipped at the 6-$\sigma$ level to isolate the two regions with potential emission.
Figure \ref{fig:specext} shows the spectra for both the potential northern and western extensions of
emission, multiplied by a factor of ten to aid in comparing to the spectrum of Leo T.
The emission from both of the potential extensions shows the same behavior: the emission
is only seen at velocities below $\sim$45 \kms, and it shows a steady increase in flux level to lower velocities
until the lowest velocity used in making the total \hi\ map. This is strong evidence that this gas is not associated
with Leo T but instead is Galactic \hi\ emission coming in at slightly higher velocities than the bulk of the Galactic
gas in this direction.



\section{Discussion}\label{sec:discuss}

Leo T is a galaxy on the edge of formation --
in addition to its low stellar and gas masses,
it is barely forming stars.
Over the last $\sim$8 Gyr, Leo T has formed stars at a low but
steady average rate 
of $\sim 5 \times 10^{-5}$ \msun\ yr$^{-1}$.
However, 25 Myr ago its star formation rate (SFR) dropped to $< 10^{-5}$ \msun\ yr$^{-1}$
\citepads{2012ApJ...748...88W}. 
While this could be a sign of the cessation of star formation in Leo T,
more likely it is a combination of the low star formation rate and stochastic
nature of star formation resulting in no massive stars formed in the last 25 Myr.

In contrast to the low rate of star formation, Leo T has a large amount
of its \hi\ mass in a CNM component. Naive pictures of how star formation proceeds in
low-mass dwarf galaxies,
with the CNM as a rough tracer of molecular hydrogen \citepads[e.g.,][]{2009ApJ...693..216K},
 would suggest that the presence of a large amount of CNM
would lead to enhanced star formation.
Below we discuss further the CNM content of Leo T, comparing to similar observations
of other dwarf galaxies. We then discuss the environment of Leo T
and postulate
that the CNM may be recently formed as a result of interaction
with the circumgalactic medium.

\subsection{The CNM and star formation in Leo T}

In Section \ref{sec:gaussdecomp},
we decomposed the \hi\ in Leo T into the CNM and WNM components to find that
 Leo T has almost 10\% of its total \hi\ mass in a CNM component.
In comparison, Leo P,
a similarly gas-rich low-mass galaxy, only has $\sim$1\% of its \hi\ mass in a CNM component
\citepads{2014AJ....148...35B}.
Indeed, while Leo P has twice the \hi\ mass of Leo T,
Leo T has $\sim$4$\times$ the amount of CNM as Leo P.
Yet, the current star formation rate in Leo P, $4.3 \times 10^{-5}$ \msun\ yr$^{-1}$
\citepads{2015ApJ...812..158M}, is higher than the current star formation rate of Leo T.

Comparing more broadly to the sample of nearby, low-mass
galaxies from \citetads{2012ApJ...757...84W}, we also find that Leo T appears to have
an unusually large amount of CNM. \citetads{2012ApJ...757...84W} find that the detected
{ fraction of a CNM component} is typically $< 5\%$ for the investigated lines-of-sight.
Our value of 10\% includes the total \hi\ content of Leo T;
if we consider only investigated lines of sight (that is, the regions of Leo T
that met the S/N requirement for the decomposition), the {fraction of \hi\ in a CNM 
component rises} to  $\sim$20\%.
At the highest column densities, the CNM dominates the WNM;
within the total \hi\ level of $4 \times 10^{20}$ atoms cm$^{-2}$, more than 50\%
of the flux comes from the CNM component at each spatial location.
This is in contrast to the galaxies in \citetads{2012ApJ...757...84W}  where typically only
$\sim$20\% of the line-of-sight contribution comes from the CNM component.
However, the dwarf galaxies of \citetads{2012ApJ...757...84W} show significantly
more star formation, controlled for their higher mass. 
While the star formation efficiency (SFE, star formation per unit gas mass)
for Leo T is $<2.4 \times 10^{-11}$, the SFE for the galaxies
in the work of \citetads{2012ApJ...757...84W} is larger than that, with about half the sample having
four times greater SFE (see their Figure 13). The difference is even more stark when comparing the SFE of the cold gas alone
{(star formation per unit gas mass in a CNM component)}.
Then, more than half of the \citetads{2012ApJ...757...84W} sample has a cold gas SFE 
{that is} ten times larger than
that of Leo T.

In addition to the fact that Leo T has more CNM than other dwarf galaxies,
its distribution of CNM is also very different. The CNM in Leo T
is always present where the total \hi\ column is above $3 \times 10^{20}$ atoms cm$^{-2}$.
In contrast, the dwarf galaxies in the \citetads{2012ApJ...757...84W}
typically only have a CNM component where the total \hi\ column density is above
the $10^{21}$ atoms cm$^{-2}$ level. Indeed, this latter value
is consistent with theoretical predictions for the transition to molecular
hydrogen \citepads[e.g.,][]{2004ApJ...609..667S,2009ApJ...693..216K}.
The physical resolution of our observations of Leo T is about twice that 
used in the \citetads{2012ApJ...757...84W} work ($117\times32$ pc vs. 200 pc). Thus, we would expect
to be able to detect more high column density emission, and so the fact that
{Leo T has a CNM component} at lower column densities than typically seen in dwarf galaxies
is not an observational artifact.

It is clear that Leo T has an unusual CNM component compared
to other gas-rich dwarf galaxies.
Its CNM comprises a larger fraction of the total \hi\ mass, is a larger
fractional component along any given line-of-sight, and is located
at sites of lower total \hi\ column density.
{We have verified that these results are robust to resolution and sensitivity effects. }
Given the large {amount of mass in a CNM component}, it is initially surprising that Leo T is not
forming stars more vigorously. 
However, its peak column density of \hi\ is only $\sim$$5\times 10^{20}$ atoms cm$^{-2}$,
well below the empirical canonical threshold for star formation
\citepads{1987NASCP2466..263S}. 
The likely explanation is that the state of the gas, {including the metallicity}, encourages
the formation of a large CNM component but not molecular hydrogen, 
the direct precursor to star formation.
{Alternatively, due to the stochastic nature of star formation at these low masses,
it could be that we happen to be observing Leo T at a specific phase where it has a large
CNM component but little to no massive star formation. }

\subsection{The environment of Leo T}\label{sec:env}

In addition to its unusual CNM component, Leo T is also an atypical gas-rich
dwarf galaxy {because it is in close proximity to a massive galaxy;}
at a distance of only 420 kpc, Leo T is close to the virial radius of the Milky Way.
Thus, we may ask if the {large CNM component} is a relatively new phenomenon,
related to the environment of Leo T. 
While the \hi\ extent of Leo T is almost completely {circular}, with no extensions of material,
there are some subtle signs of disturbance. The main \hi\ peak, including the location of the CNM component,
is offset to the south from the optical (and \hi) center of the galaxy.
In addition, the outermost \hi\ contours on the western side are more compressed than elsewhere
in the galaxy; this is also seen in Figure \ref{fig:radialprofile} where the western column density
profile falls off more steeply than in any other direction.
The compressed \hi\ contours could be a truncation of the \hi\ disk
from interaction with the circumgalactic medium of the Milky Way 
\citepads[e.g., as seen for the Large Magellanic Cloud;][]{2015ApJ...815...77S}.
{Internal compression of the gas from ram pressure could cause the offset \hi\ peak,
and also contribute to the presence of a relatively large CNM component}
\citepads[e.g.,][]{2003MNRAS.345.1329M,2006MNRAS.369.1021M}.
Given that Leo T is gas-rich, and the known morphological segregation in the Local Group
implies that gas is removed from dwarf galaxies that interact closely with the
 Milky Way or Andromeda galaxy \citepads{2014ApJ...795L...5S},
these hints of interaction are consistent with the idea that Leo T
is infalling onto the Galaxy for a first time \citepads{2012MNRAS.425..231R}.
Thus, {one possible explanation for the relatively large amount of CNM may be 
that it is a recent transient phenomenon}, related to the infall
of Leo T onto the Milk Way.
{In this scenario, the observed drop in star formation rate in Leo T could be real,
due to changes in the internal gas structure, and not stochastic sampling of a steady star formation rate.
Future simulations and modeling to understand if ram pressure can produce
a large CNM component, especially with an offset spatial position and  velocity, would shed light on this
potential scenario.
}





\section{Conclusions}\label{sec:concl}

We present deep WSRT \hi\ observations of Leo T with a spatial resolution of  117$\times$32 pc
and a spectral resolution of 0.52 \kms. Our main conclusions are as follows:
\begin{itemize}
\item We find a total \hi\ mass of $4.1 \times 10^5$ \msun\ for a distance of 420 kpc. This is an increase
of $\sim$50\% compare to the previously commonly accepted value.
\item At a column density level of $2.7 \times 10^{19}$ atoms cm$^{-2}$, the \hi\ radius is 400 pc.
We can integrate to trace the \hi\ emission to the $6 \times 10^{18}$ atoms cm$^{-2}$ level at 
a radius of 500 pc.
\item The \hi\ distribution of Leo T is very regular, with no evidence for emission extending
from the main \hi\ disk. There are, however, more subtle signs of the interaction of Leo T
with the circumgalactic medium, including a truncation of the \hi\ emission on the western edge and the offset
of the peak \hi\ emission to the south from the optical center.
\item A Gaussian decomposition of the \hi\ emission reveals that  
Leo T has a significant amount of its \hi\ mass in the CNM phase, $\sim$10\%,
with the potential to contribute more than 50\% of the flux along the lines-of-site where
it is located. Despite
this large amount of CNM compared to other gas-rich dwarf galaxies, Leo T is not currently forming stars
and has a low star formation efficiency compared to other dwarf galaxies.
{One possible explanation is that the} the formation of the CNM phase is relatively recent,
tied to the infall of Leo T into the Milky Way and interaction with the circumgalactic medium.
{Alternatively, it could be that due to the stochastic nature of star formation in low mass galaxies,
we are observing Leo T at a phase where it happens to have a large CNM component with little to no massive
star formation.}
\end{itemize}



\begin{acknowledgements}
{We wish to thank the anonymous referee for useful comments that improved the quality of this manuscript.
We also wish to thank N. Giese, P. Serra, and T. Westmeier for help with installing and running SoFiA.}
The Westerbork Synthesis Radio Telescope is operated by ASTRON, the Netherlands Institute for Radio Astronomy, with support from the Netherlands Foundation for Scientific Research (NWO).
This work is part of the research programme HuDaGa with project number
TOP1EW.14.105, which is financed by the Netherlands Organisation for
Scientific Research (NWO).
EAKA is supported by the WISE research programme, which is financed by the Netherlands Organisation for Scientific Research (NWO).
This research made use of APLpy, an open-source plotting package for Python hosted at http://aplpy.github.com; Astropy, a community-developed core Python package for Astronomy (Astropy Collaboration, 2013);  the NASA/IPAC Extragalactic Database (NED) which is operated by the Jet Propulsion Laboratory, California Institute of Technology, under contract with the National Aeronautics and Space Administration; and NASA's Astrophysics Data System.

\end{acknowledgements}



\bibliographystyle{aa}
\bibliography{refs}



\appendix
\section{The \hi\ line flux of Leo T}\label{app:flux}

{Our spectrum of Leo T  (Figure \ref{fig:spec})
reveals 
a total line flux of 
$9.6 \pm 0.9$ Jy \kms, obtained from fitting two Gaussian components to account
for confusion with Galactic \hi.
This is significantly} higher than the commonly accepted value for Leo T of
6.7 Jy \kms\ \citepads{2008MNRAS.384..535R}. In this Appendix, we undertake a detailed determination of the line flux of Leo T, 
with a comparison to literature values for the \hi\ flux of Leo T. 
Table \ref{tab:fluxvals} compiles our line flux values, {always
obtained from fitting two Gaussians to the spectrum}, in addition to literature values.
We report a final line flux value for Leo T of 9.9 $\pm$ 1.0 Jy \kms,
{based on spatially smoothing the total intensity map to define
the extent for extracting a spectrum}.

We find that
for
shallow interferometric data, the method used to select emission is critical when much
of the \hi\ mass is in low column density emission. 
{For sources where the velocity does not change significantly as a function of spatial position,
creating a total intensity \hi\ map by collapsing all channels with emission (without any masking),
aids significantly in flux recovery. Using this methodology, the full line flux of Leo T can be recovered even
in shallow data.
}

\subsection{Determining the extent of emission}
First, we carefully consider how we derive the line flux from the WSRT observations presented in this work.
In addition to the spectrum from a 3-$\sigma$ clip (Figure \ref{fig:spec}), 
we also use a more expansive mask to derive another spectrum of Leo T,
to understand the robustness of our line flux value.
We smoothed the integrated \hi\ map to 160\arcsec\ and clipped
at the 10-$\sigma$ level. The second mask has a larger spatial extent and should include 
more low level emission than the direct 3-$\sigma$ clip. The angular smoothing is the same as used for 
defining the spatial extent of Leo T for creating a clean mask (Section \ref{sec:data}). 
In this case, the clip level
is high in order to avoid including the northern extension of emission,
so we may still be missing some faint emission.
The direct line flux from this more expansive mask is 9.7 Jy \kms,
a slight increase from the direct clip mask ({9.4 Jy \kms}) as we expect.

We also fit single and two Gaussian components to the spectrum from the smoothed mask
to derive an integrated line flux that includes extrapolating
 into the region of Galactic \hi\ confusion.
The two Gaussian component fit is significantly better than a single component,
and we find a higher line flux of 9.9 Jy \kms.
This spectrum and fit are shown in Figure \ref{fig:specflux}, with the
3-$\sigma$ spectrum from Figure \ref{fig:spec} for reference.
We adopt $9.9 \pm 1.0$ Jy \kms\ as our line flux value for Leo T,
with the 10\% uncertainty reflecting the accuracy of our calibration and
other systematics, including exactly how Leo T emission is selected.
The small difference between this value and that based on a direct 3-$\sigma$ clip implies that only 
$\sim$3\% of the total \hi\ flux comes from emission below the $1.6 \times 10^{19}$ 
atoms cm$^{-2}$ level ({at a resolution of 57.3\arcsec $\times$ 15.7\arcsec}).
{As shown in Figure \ref{fig:mom0}, this contour level has an extent of $\sim$3.3\arcmin.
The full \hi\ extent is traced to $\sim$4.2\arcmin\ (Section \ref{sec:hidist}) at the $6 \times 10^{18}$
atoms cm$^{-2}$ level. This is almost a 30\% increase in size while only accounting for 3\% of the total flux.
Thus, it is difficult to know when a galaxy edge is found
if relying on a methodology that uses a flux growth curve
as deeper integration will reveal low column density emission at further extents that does not contribute significantly
to the total line flux.
}

\begin{figure}
\centering
\includegraphics[keepaspectratio,width=\linewidth]{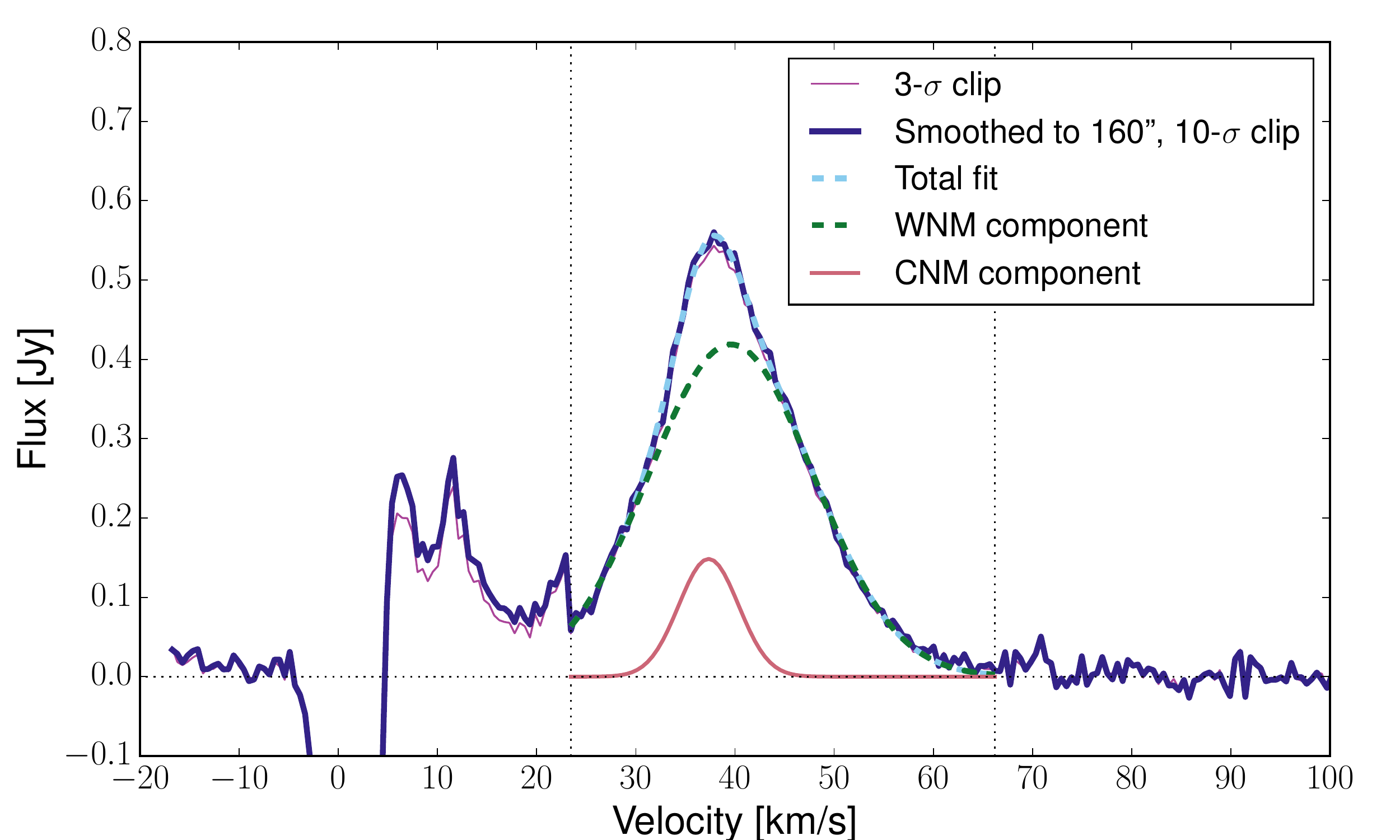}
\caption{Spectrum {based on smoothing the total intensity map
to define the spatial extent.}
The spectrum from
the directly clipped mask (Figure \ref{fig:spec}) is shown for comparison. 
A two component Gaussian fit used to integrate the full line flux of Leo T is also
shown.}
\label{fig:specflux}
\end{figure}

\begin{table*}
\centering
\caption{Line flux values for Leo T from global fits to spectra}
\label{tab:fluxvals}
\begin{tabular}{lllll}
\hline \hline
Observation & $S_{int}$ & $S_{CNM}$ & $S_{WNM}$ & Reference\\
   & Jy \kms & Jy \kms & Jy \kms & \\
\hline
Deep WSRT, 3-$\sigma$ & $9.6 \pm 1.0$ & $1.1 \pm 0.1 $ & $8.4 \pm 0.8$ & 1 \\
Deep WSRT, 160\arcsec, 10-$\sigma$ & $9.9 \pm 1.0$ & $1.1 \pm 0.2$ & $8.7 \pm 0.9$& 1\\
Shallow WSRT, 160\arcsec, 4-$\sigma$ & $9.9 \pm 1.0$ & $0.80 \pm 0.18 $ & $9.1 \pm 0.9$ & 1\\
Deep WSRT, {channel mask} & ${8.7 \pm 0.9}$  & ${0.69 \pm 0.11} $ & ${8.0 \pm 0.8}$ & 1 \\
{Shallow WSRT, channel mask} & ${6.1 \pm 0.6}$ & ${0.0 \pm 0.0}$ & ${6.1 \pm 0.6}$ & 1\\
Original WSRT & 6.7  & .. & .. & 2\\
HIPASS & 4.9 & .. & .. & 3\\
HIPASS & 10. & .. & .. & 4 \\
GALFA & 12. & .. & .. & 4 \\
ALFALFA & 11.4 & .. & .. & 5,6 \\
\hline
\end{tabular}
\tablefoot{
1: This work, 2: \citetads{2008MNRAS.384..535R}, 3: \citetads{2007ApJ...656L..13I},
4: \citetads{2009ApJ...696..385G}, 5: \citetads{2011AJ....142..170H}, 6: The $\alpha$.70 catalog is publicly
available at http://egg.astro.cornell.edu/alfalfa/data/index.php
}
\end{table*}

\subsection{Comparison to shallow WSRT data}

Our reported line flux value is $\sim$50\% higher
than that of \citetads{2008MNRAS.384..535R},
derived from a single 12-hr WSRT observation.
While the discrepancy in flux between the deep observations presented here and the observations
of \citetads{2008MNRAS.384..535R} could be understood as a difference in sensitivity, the approach
for isolating the emission from Leo T also varies between our work and that of \citetads{2008MNRAS.384..535R}.
We used an expansive {frequency-independent clean mask}, so that at a spatial location where there is emission, we cleaned that location
for all channels that contain any source emission. 
{This allowed us to then create our total intensity map by collapsing all channels with emission without
any preselection or masking.}
In contrast, \citetads{2008MNRAS.384..535R} take
a more traditional approach,
{for both their cleaning and creation of a total intensity \hi\ map, where they identify}
the emission in each channel individually by smoothing and clipping on a threshold. 
This approach is appropriate for large, massive galaxies where the rotation velocity is large so that emission at a given velocity is only in a small spatial region of the galaxy. 
In {most of} these systems the majority of the \hi\ mass is also in high column density material that is easy to detect.
However, for low mass objects with little velocity structure and a large fraction of their \hi\ in low column density material,
this approach may exclude some material from being included in the defined regions of the galaxy.
Indeed,  looking at the comparison of our integrated \hi\ map to that of \citetads[][Figure \ref{fig:momcompare}]{2008MNRAS.384..535R}, contours at the same level have a smaller spatial extent in the map of
\citetads{2008MNRAS.384..535R}, hinting that missing emission from that map may be the explanation
for the line flux discrepancy.

\begin{figure}
\centering
\includegraphics[keepaspectratio,width=\linewidth,clip=true,trim=1cm 3cm 0cm 1cm]{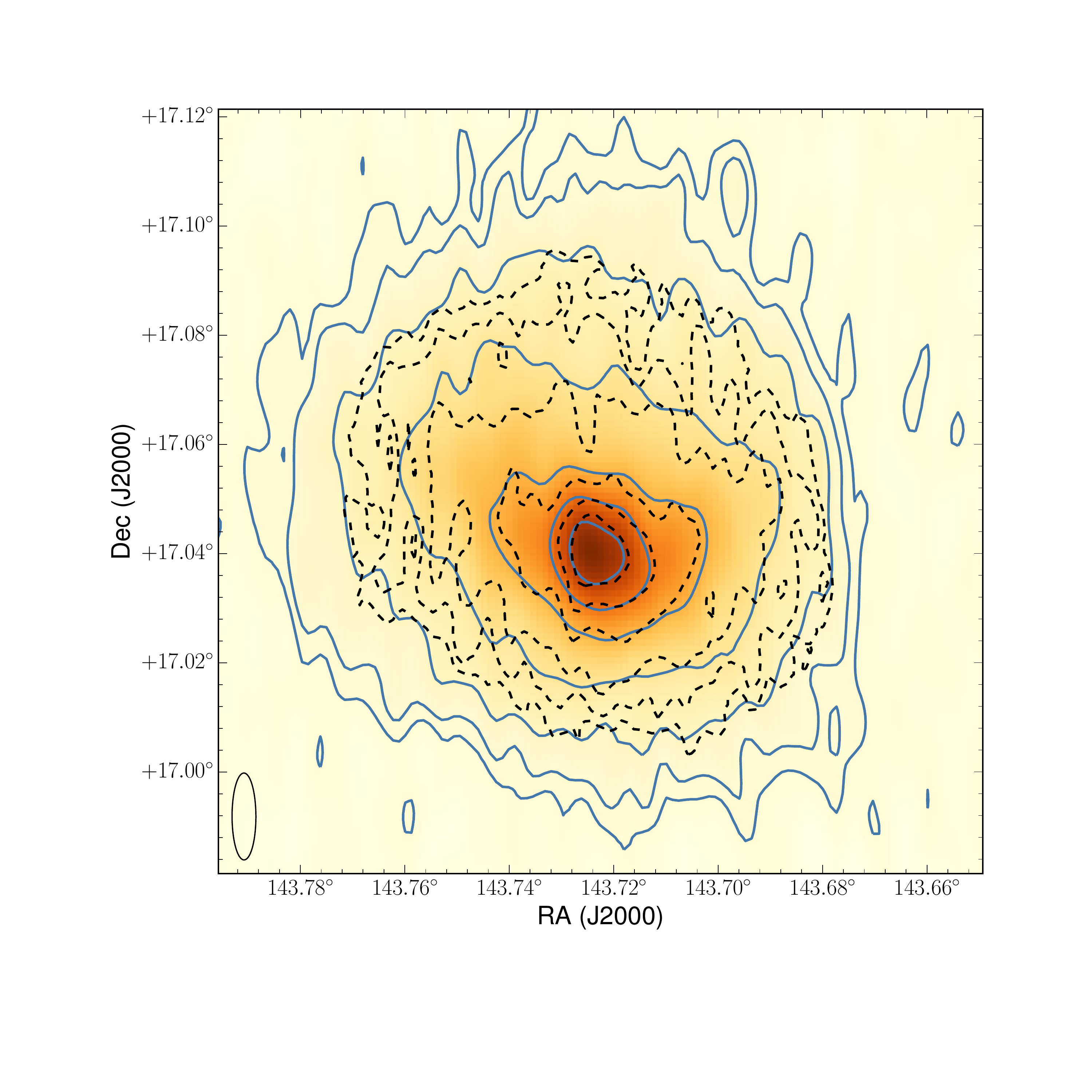}
\caption{Comparison of the original, shallow WSRT moment zero map \citepads[color-scale and dotted contours;][]{2008MNRAS.384..535R})
with the deeper WSRT observations (solid contours). Contour levels are at [1.5, 2.5, 5, 10, 20, 30, 40] $\times 10^{19}$ atoms cm$^{-2}$
as in Figure \ref{fig:mom0}.}
\label{fig:momcompare}
\end{figure}

To test this, we imaged a single WSRT observation of Leo T but using the same clean mask
derived for the full dataset (described in detail in Section \ref{sec:data}).
{We combined all channels identified as having emission, without any clipping or selection,
to create the total intensity \hi\ map. }
Figure \ref{fig:mom0_d1only} shows the comparison of the column density map of this single dataset
to the deep observations presented here. While this data is not as deep, the column density contours
have the same spatial distribution as the deep observations presented here, in
 contrast to the previous analysis of the shallow WSRT observations.

\begin{figure}
\centering
\includegraphics[keepaspectratio,width=\linewidth,clip=true,trim=1cm 3cm 0cm 1cm]{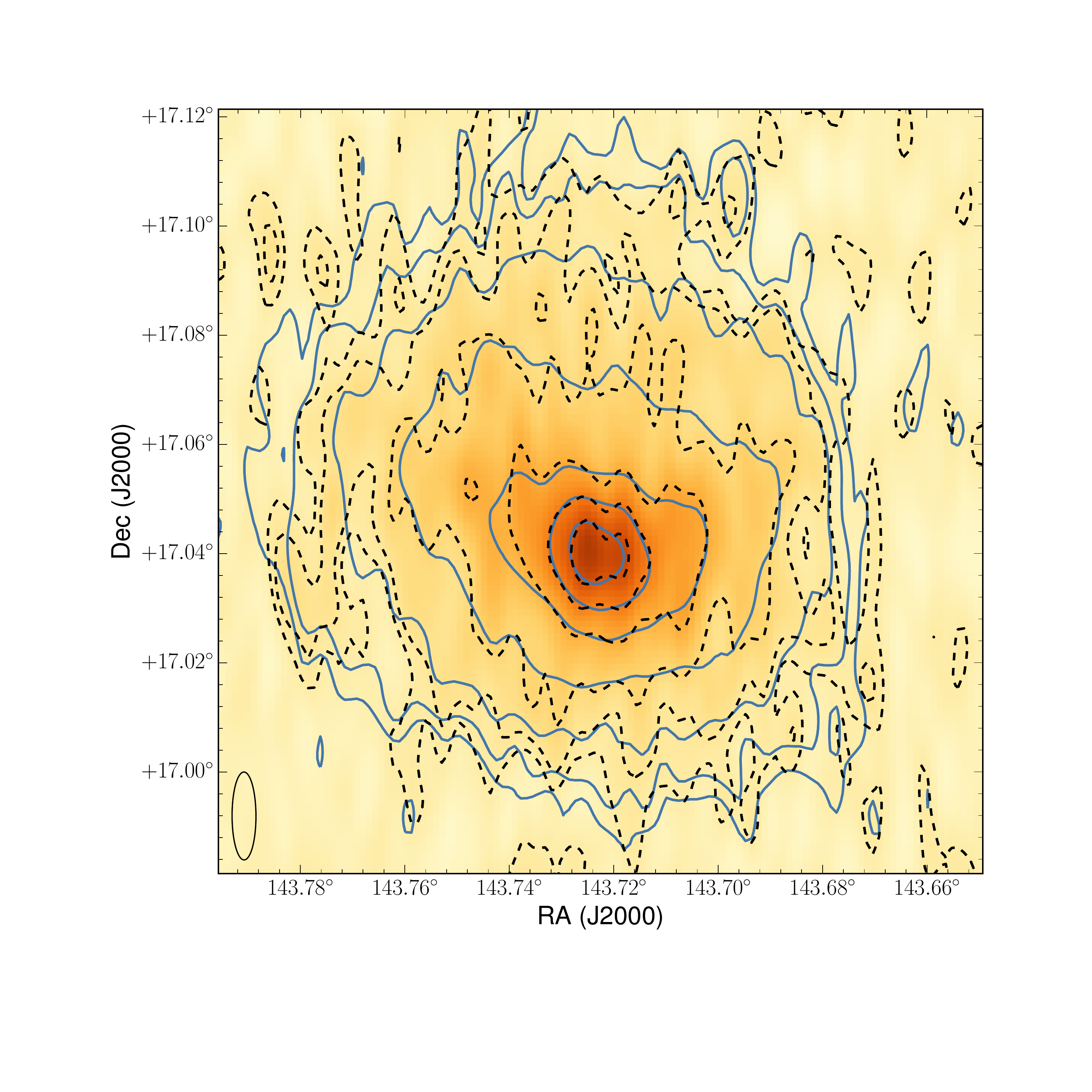}
\caption{Comparison of shallow and deep WSRT data with the same masking strategy. The color map and dotted contours are {a} single WSRT observation of Leo T but with the {frequency-independent mask used}
 in this work. The black contours are from the full deep WSRT observations. The levels are
[2.5, 5, 10, 20, 30, 40] $\times 10^{19}$ atoms cm$^{-2}$. The lowest contour is slightly less than 5-$\sigma$ for the deep data
and 1.6-$\sigma$ for the shallow data.}
\label{fig:mom0_d1only}
\end{figure}

In order to test how the flux recovery is improved by using this 
{alternative approach to creating a total intensity \hi\ map without masking,}
we derived the spectra shown in Figure \ref{fig:specd1} in a similar manner to the spectra found for the deep data.
We smoothed the total intensity \hi\ map to 160\arcsec\ angular resolution and then clipped at the 4-$\sigma$ level.
This is a much more stringent clip than used in the deep data; this is because the presence of the contaminating
Galactic \hi\ foreground is suppressed in the shallower data and so we could create a deeper mask without including extensions
that are potentially Galactic. While the spectrum for the single dataset is much noisier, it matches the spectra
from the deeper observations exactly. We fit a two Gaussian component to this spectrum to find a line flux of
9.9 Jy \kms, exactly matching that of the deep WSRT data. However, we note that the decomposition of this spectrum
into Gaussian components is different compared to the deep data, {although consistent within the errors}.
This demonstrates that, even with shallow data, with careful masking and determination of source extent,
the full emission of low surface brightness objects can be recovered.
It also indicates that the lowest column density material at the edges of galaxies does not contribute
significantly to the total line flux. Thus it is difficult to know when the edge of a galaxy is found as further
integration reveals more low column density emission at further extents without impacting the total line flux.

To again illustrate the importance of the determination of source extent, we derived another spectrum from the shallow data,
also shown in Figure \ref{fig:specd1}.
In this case, we clipped directly at the 3-$\sigma$ level on the total intensity \hi\ map ({made without
any channel-based masking}),
without any spatial smoothing. 
The flux recovered in this spectrum (8.4 Jy \kms) is slightly suppressed compared to the 
deep WSRT spectrum. 
In this case, the spatial extent of the mask is smaller, resulting in less emission from Leo T being included.

\begin{figure}
\centering
\includegraphics[keepaspectratio,width=\linewidth]{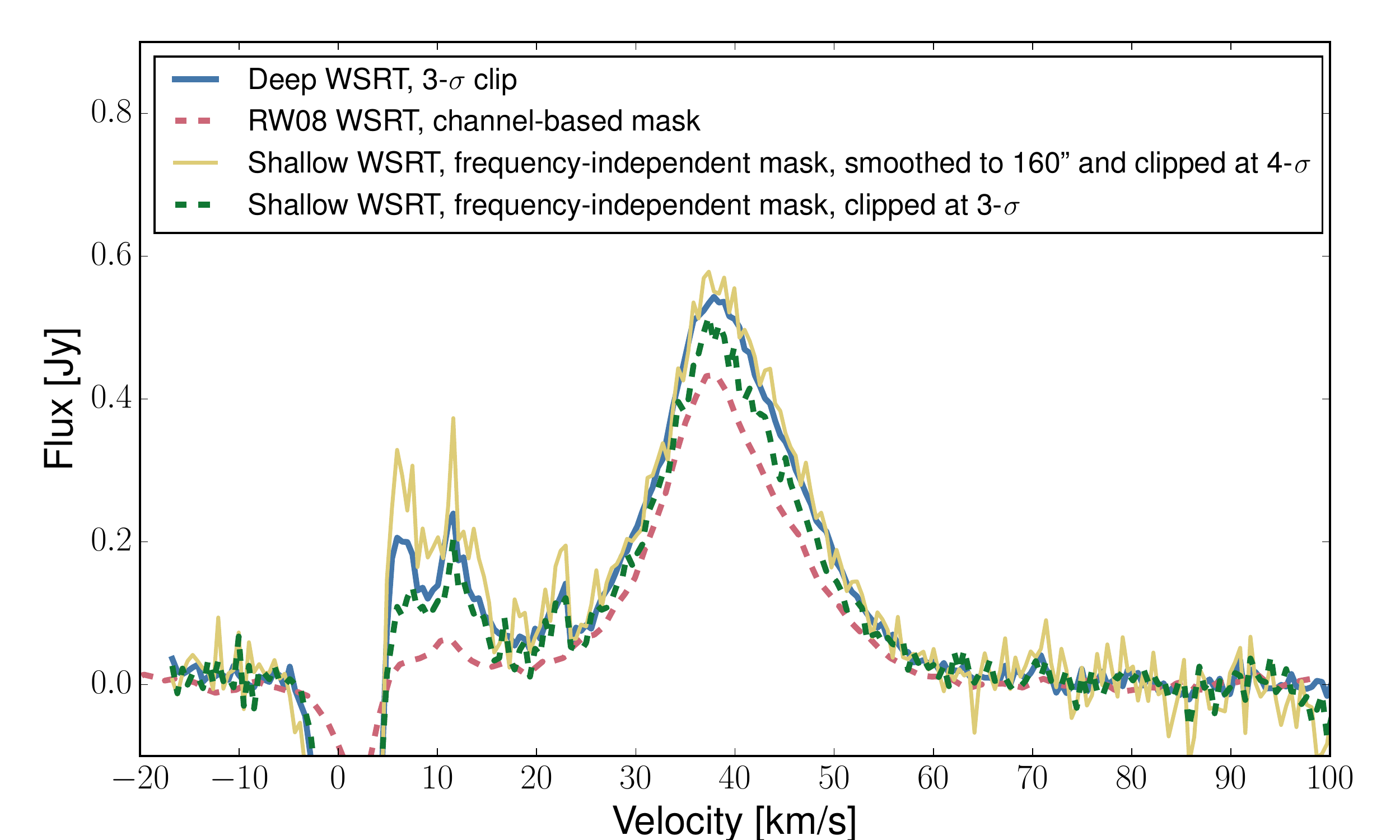}
\caption{Comparison of spectra of Leo T from the deep observations and single shallow observations with different masking
strategies, including {frequency-independent clean masks}.}
\label{fig:specd1}
\end{figure}

\subsection{Comparison to channel-based masking}

{
To demonstrate the importance of frequency-independent versus channel-based masking in different S/N regimes,
we used {\tt SoFiA} to produce channel-based masks for both the deep and shallow WSRT data. 
For both datasets, we used the same key parameters.
We set: threshold $=$ 4, merging radius $=$ 1 for all dimensions,
reliability threshold $=$0.9, and kernel scale$=$0.5.
The most important parameters are the initial smoothing of the data cube and the kernels
used for smoothing during the source finding.
We used an input smoothing of three spatial pixels and one channel, and set the kernels for the source finding to be three and six spatial pixels and three channels in the velocity dimension.
These parameters were chosen to match what is traditionally used for creating channel-based source masks.
}

{
Figure \ref{fig:specsofia} shows the resulting two spectra for both the deep and shallow data compared to the frequency-independent
mask for the deep data. The result is that a channel-based mask for the shallow data misses a large amount of emission; only around 60\%
of the the flux is recovered (see Table \ref{tab:fluxvals}).
However, for the deep data presented in this work, the channel-based mask recovers a majority of the emission, $\sim$90\% of the
total line flux. This is because with the increased sensitivity, the channel-based identification of the emission includes
lower column density material missed in the shallow data. Importantly, the deep data with a channel-based mask still has a reasonable 
(and consistent within the errors) decomposition into CNM and WNM components based on the global spectrum, while the shallow
data with a channel-based mask shows no evidence for a CNM component in the global spectrum.
}

{
We also note that by adjusting the parameters of SoFiA the full line flux of Leo T can be recovered, even with the shallow data.
Implementing a strong velocity smoothing to the input data cube or using large velocity kernels for the source finding effectively
mimics our strategy of defining the source extent without channel-based masking. 
}

\begin{figure}
\centering
\includegraphics[keepaspectratio,width=\linewidth]{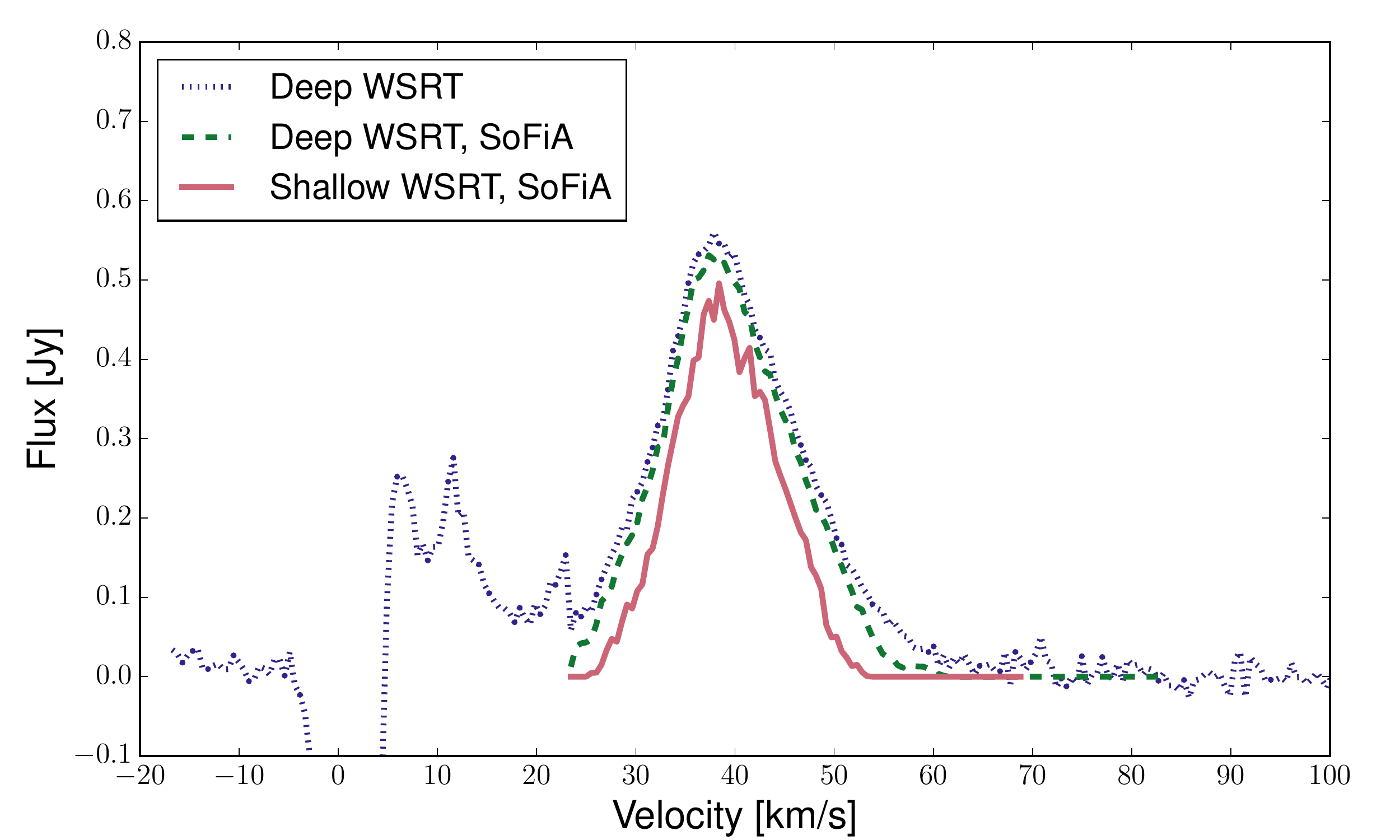}
\caption{Spectra of both deep and shallow Leo T observations {from channel-based masks found
with SoFiA.}}
\label{fig:specsofia}
\end{figure}

\subsection{Comparison to literature}

The line flux we find for Leo T is 50\% higher than the commonly used value of 6.7 Jy \kms\ reported by  \citetads{2008MNRAS.384..535R}. Their value is consistent with that initially found based on HIPASS data \citepads{2007ApJ...656L..13I}.
However, \citetads{2009ApJ...696..385G} later looked at the HIPASS data to find a \hi\ flux for Leo T of 10 Jy \kms.
HIPASS has a spatial resolution of 15.5\arcmin\ and a velocity resolution of $\sim$26 \kms, so it is difficult to disentangle Leo T from the foreground Galactic \hi\ emission in those data \citepads{2004MNRAS.350.1195M}. The ALFALFA \hi\ survey has a spatial resolution of 3.5\arcmin\ and a velocity resolution of $\sim$11 \kms, making it easier to separate Leo T spatially and kinematically from the Galactic foreground \citepads{2011AJ....142..170H}. The
footprint of the 70\% complete survey\footnote{The 70\% ALFALFA survey is publicly available at http://egg.astro.cornell.edu/alfalfa/data/index.php} contains Leo T, and a higher line flux of 11.4 Jy \kms\ is also found in that data.
In Figure \ref{fig:speccompare}, we compare the ALFALFA spectrum to our WSRT spectrum and that from \citetads{2008MNRAS.384..535R}. While the peak flux of the ALFALFA spectrum is lower than our spectrum, the broader velocity extent and total line flux agree well with our data. There is some blending of Leo T with the Galactic emission in the ALFALFA spectrum, potentially explaining the slightly higher line flux for the ALFALFA data. \citetads{2009ApJ...696..385G} also reported the \hi\ mass for Leo T from GALFA data, which has the same spatial resolution as ALFALFA but a higher spectral resolution of 0.74 \kms; their corresponding line flux value is $\sim$12 Jy \kms. 

The single-dish \hi\ line flux values for Leo T are higher than we find. However, it is difficult to spatially isolate Leo T emission from Galactic \hi\ emission in the WSRT data, and the problem is exasperated for single-dish telescopes with their poorer angular resolution. Thus, these line flux values likely included some Galactic \hi\ contribution. On the other hand, it is possible that we are missing from Leo T due to restricting our clip to avoid Galactic \hi\ emission. However, we note that our line flux based on smoothing the moment zero map only increased the total line flux by $\sim$3\%, even thought the total area considered increased by almost 30\%. Thus we are confident that the total line flux reported here accurately represents the total emission of Leo T, modulo remaining systematic uncertainties resulting from blending with Galactic \hi\ emission.

\begin{figure}
\centering
\includegraphics[keepaspectratio,width=\linewidth]{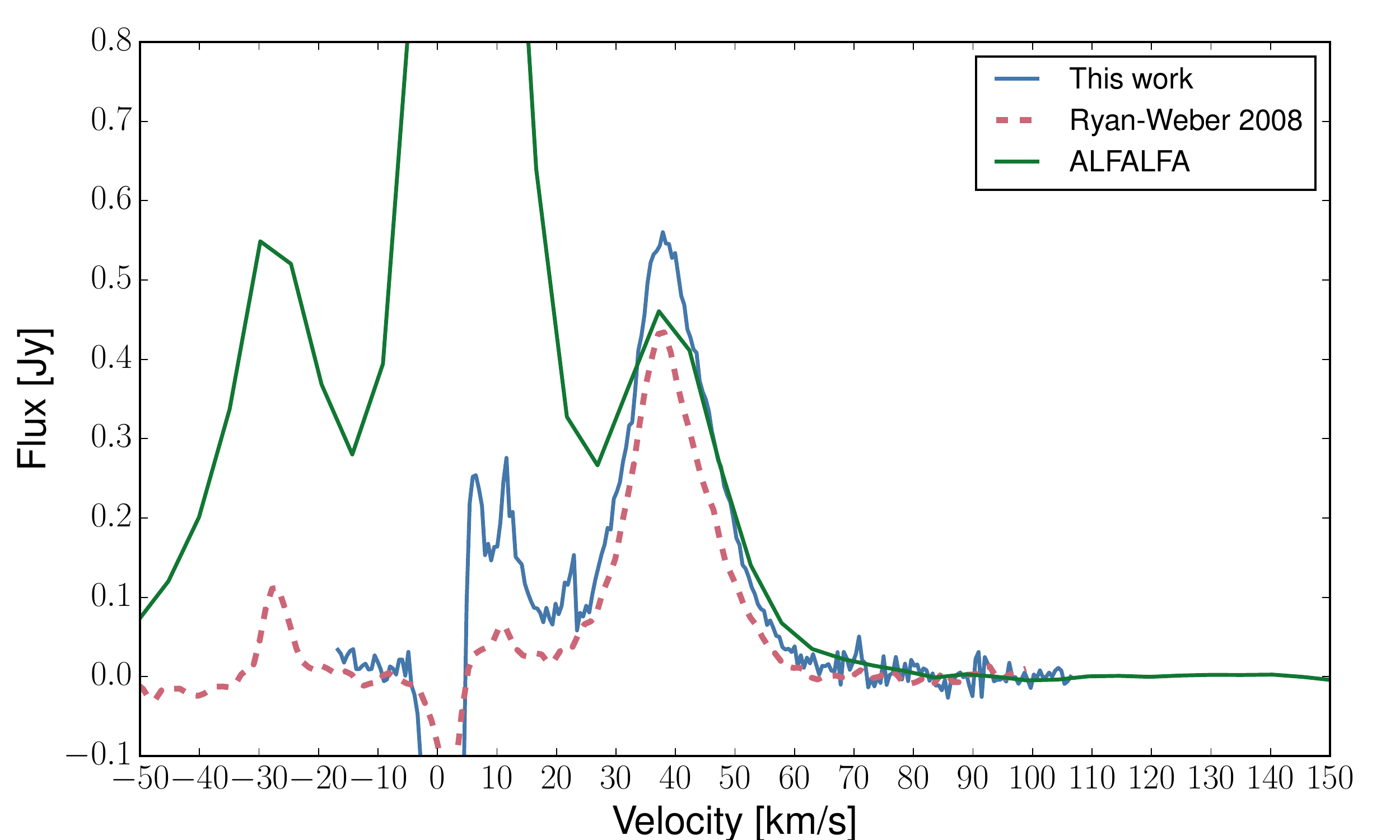}
\caption{WSRT spectra of this work, with the ALFALFA and previous shallow WSRT spectra
for comparison.}
\label{fig:speccompare}
\end{figure}

\end{document}